\input harvmac
\input tables
\input epsf
\def\figin{\epsfcheck\figin}\def\figins{\epsfcheck\figins}
\def\epsfcheck{\ifx\epsfbox\UnDeFiNeD
\message{(NO epsf.tex, FIGURES WILL BE IGNORED)}
\gdef\figin##1{\vskip2in}\gdef\figins##1{\hskip.5in}
\else\message{(FIGURES WILL BE INCLUDED)}%
\gdef\figin##1{##1}\gdef\figins##1{##1}\fi}
\def\DefWarn#1{}
\def\figinsert{\goodbreak\midinsert}
\def\ifig#1#2#3{\DefWarn#1\xdef#1{fig.~\the\figno}
\writedef{#1\leftbracket fig.\noexpand~\the\figno}%
\figinsert\figin{\centerline{#3}}\medskip\centerline{\vbox{\baselineskip12pt
\advance\hsize by -1truein\noindent\footnotefont{\bf Fig.~\the\figno:} #2}}
\bigskip\endinsert\global\advance\figno by1}

%
\def\np#1#2#3{Nucl. Phys. B{#1} (#2) #3}

\def\tilde{\widetilde}

\def\Tr{{\rm Tr ~}}

\font\zfont = cmss10 
\def\ZZ{\hbox{\zfont Z\kern-.4emZ}}

\def\nc{{N_{c}}}

\def\nfl{{N_{f}}}\def\nfr{{N'_{f}}}
\def\ncl{{\nc}}\def\ncr{{N'_{c}}}

\def\ncld{{\tilde{N}_c}}\def\ncrd{{\tilde{N}'_c}}

%
%
\def\ts{F}
\def\Ql{Q}\def\Qr{Q'}
\def\Pl{M}\def\Pr{M'}
\def\Mm{P}\def\Mmt{\tilde{\Mm}}
\def\Qlt{\tilde{\Ql}}\def\Qrt{\tilde{\Qr}}
\def\tst{\tilde{F}}
%
%
\def\tsD{f}\def\tstD{\tilde\tsD}
\def\ql{q}\def\qr{\ql'}
\def\qlt{\tilde\ql}\def\qrt{\tilde\qr}

\def\journal#1&#2(#3){\unskip, \sl #1\ \bf #2 \rm(19#3) }
\def\andjournal#1&#2(#3){\sl #1~\bf #2 \rm (19#3) }

\def\tst#1{{\textstyle #1}}
\def\frac#1#2{{#1\over#2}}

\def\inbar{\,\vrule height1.5ex width.4pt depth0pt}
\def\IC{\relax\hbox{$\inbar\kern-.3em{\rm C}$}}
\def\IR{\relax{\rm I\kern-.18em R}}
\def\IP{\relax{\rm I\kern-.18em P}}

%
%
\def\np#1#2#3{Nucl. Phys. {\bf B#1} (#2) #3}

\catcode`\@=11
\def\slash#1{\mathord{\mathpalette\c@ncel{#1}}}
\overfullrule=0pt

\def\ZZ{{\cal Z}}

\def\underrel#1\over#2{\mathrel{\mathop{\kern\z@#1}\limits_{#2}}}

\catcode`\@=12


%


\def\nsp{{NS$^\prime$}}

\def\tst{\tilde{F}}


\lref\nati{N. Seiberg, {\it Electric-Magnetic Duality in Supersymmetric 
Non-Abelian
Gauge Theories}, \np{435}{1995}{129}, hep-th/9411149.}

\lref\kuts{D. Kutasov, {\it A Comment on Duality in N=1
Supersymmetric Non-Abelian Gauge Theories}, EFI--95--11,
hep-th/9503086.}

\lref\kutsch{ D. Kutasov and A. Schwimmer, {\it On Duality in
Supersymmetric Yang-Mills Theory}, EFI--95--20, WIS/4/95,
hep-th/9505004.}

\lref\Ahsonyank{O. Aharony, J. Sonnenschein, and S. Yankielowicz,
{\it Flows and Duality Symmetries in N=1 Supersymmetric Gauge
Theories}, TAUP--2246--95, CERN-TH/95--91, hep-th/9504113.}

\def\kut{\refs{\kuts , \kutsch}}

\lref\witten{E. Witten, {\it Solutions Of Four-Dimensional Field Theories Via M 
Theory}, hep-th/9703166.}

\lref\brodie{J.H. Brodie, {\it Duality in Supersymmetric SU($N_c$) 
Gauge Theory with
Two Adjoint Chiral Superfields}, hep-th/9605232, 
Nucl.Phys. B478 (1996) 123-140}

\lref\bs{J. H. Brodie and M. J. Strassler, {\it Patterns of Duality in N=1 SUSY 
Gauge Theories},  PU-1658, IASSNS-HEP-96/110,hep-th/9611197.}

\lref\keg{S. Elitzur, A. Giveon, and D. Kutasov, {\it Branes and N=1 Duality in 
String Theory}, hep-th/9702014.}

\lref\ils{K. Intriligator, R. G. Leigh, and M. J. Strassler, {\it New Examples 
of Duality in Chiral and Non-Chiral
Supersymmetric Gauge Theories}, hep-th/9506148, Nucl.Phys. B456 (1995) 567.}

\lref\hw{A. Hanany and E. Witten, {\it Type IIB Superstrings, BPS Monopoles, 
And Three-Dimensional Gauge Dynamics}, IASSNS-HEP-96/121, hep-th/9611230.}

\lref\barb{J. L. F. Barbon, {\it Rotated Branes and N=1 Duality}, 
CERN-TH/97-38, hep-th/9703051.}

\lref\asp{ P. C. Argyres, M. R. Plesser, Nathan Seiberg, 
{\it The Moduli Space of N=2 SUSY QCD and 
Duality in N=1
SUSY QCD}, RU-96-07, WIS-96-1, hep-th/9603042}

\lref\berkooz{M. Berkooz, M. R. Douglas, R. G. Leigh, 
{\it Branes Intersecting at Angles}, hep-th/9606139,  
Nucl.Phys. B480 (1996) 265-278.}

\lref\branesendingI{A. A. Tseytlin,
{\it No-force condition and BPS combinations of p-branes in 11 and 10 dimensions
}, hep-th/9609212, Nucl.Phys. B487 (1997) 141.}

\lref\branesendingII{R. Argurio, F. Englert, L. Houart,
{\it Intersection Rules for p-Branes},
hep-th/9701042.}

\lref\berkeley{J. de Boer, K. Hori, H. Ooguri, Y. Oz, Z. Yin,
{\it Mirror Symmetry in Three-Dimensional Gauge Theories,
SL(2,Z) and D-Brane Moduli Spaces}, LBNL-039707, UCB-PTH-96/58, 
hep-th/9612131;
J. de Boer, K. Hori, Y. Oz, Z. Yin,
{\it Branes and Mirror Symmetry in N=2 Supersymmetric Gauge
Theories in Three Dimensions},  LBNL-039423, UCB-PTH-97/09,
hep-th/9702154.
}

\lref\ahiss{O. Aharony, A. Hanany, K. Intriligator, N. Seiberg, 
and M.J. Strassler 
{\it Aspects of N=2 Supersymmetric Gauge Theories in Three
Dimensions}, RU-97-10, IASSNS-HEP-97/18, hep-th/9703110.}

\lref\cliff{N. Evans, C. V. Johnson, and A. D. Shapere,
{\it Orientifolds, Branes, and Duality of 4D Gauge Theories}, 
BUHEP-97-11, UK/97-05, hep-th/9703210}

\lref\ovafa{H. Ooguri and C. Vafa, {\it 
Geometry of N=1 Dualities in Four Dimensions},  HUPT-97/A010, 
UCB-PTH-97/11, LBNL-40032, hep-th/9702180.}

\lref\ofer{O. Aharony, {\it IR Duality in $d=3$ $N=2$ Supersymmetric 
$USp(2N_c)$ and $U(N_c)$ Gauge Theories}, RU-97-18, hep-th/9703215.}

\lref\kar{ A. Karch, {\it Seiberg duality in three dimensions} 
 HUB-EP-97/20, hep-th/9703172.}

\lref\ah{O. Aharony and A. Hanany, unpublished.}

\Title{\vbox{\rightline{hep-th/9704043}\rightline{PU--1695 }
\rightline{IASSNS--HEP--97/35}}}
{\vbox{\centerline{Type IIA Superstrings, Chiral Symmetry, and}
\medskip
\centerline{$N=1$ 4D Gauge Theory Dualities}}}
\centerline{ John H. Brodie}
\smallskip
{\it
\centerline{Department of Physics}
\centerline{Princeton University}
\centerline{Princeton, NJ 08540, USA}}
\centerline{\tt jhbrodie@princeton.edu}
\bigskip\medskip
\centerline{Amihay Hanany}
\smallskip{\it
\centerline{School of Natural Sciences}
\centerline{Institute for Advanced Studies}
\centerline{Princeton, NJ 08540, USA}}
\centerline{\tt hanany@ias.edu}



\noindent
We study $N=1$ four dimensional gauge theories as the world volume theory
of D4-branes between NS 5-branes. 
We find constructions for a number of known field theory
dualities involving $SU(N_c)\times SU(N_c')$ groups,
coupled by matter fields $F$ in the $(N_c, \bar N_c)$ representation,
in terms of branes of type IIA string theory.
The dual gauge group follows from simply
reversing the ordering of the NS 5-branes and the D6-branes while
conserving magnetic charge on the world volume of the branes. 
We interpret many field theory phenomena such as deformation of 
the superpotential $W = \Tr (F\tilde F)^{k+1}$
in terms of the position of branes. 
By looking to D-branes for guidance,
we find new $N=1$ dualities 
involving arbitrary numbers of gauge groups.
We propose a mechanism for enhanced chiral symmetry in the brane construction
which, we conjecture, 
is associated with tensionless threebranes in six dimensions.
\Date{4/97}

\newsec{Introduction}

The world volume field theory on the branes of superstring theory has 
received much attention recently. Many new field theory results as well as 
new implications
for string theory dynamics have been obtained by
studying the low energy physics of branes in space-time.
One particular construction was introduced in \hw\ to study the dynamics of
2+1 dimensional gauge theories using Type IIB physics.
This construction has lead to applications in various dimensions and
supersymmetries
{\refs{\berkeley, \keg, \ovafa, \ahiss, \witten, \kar, \cliff, \ofer}.
A common approach in all these papers
is to stretch 4 (or 3)-branes of type IIA (or IIB) string theory 
between two NS 5-branes producing what looks to be at large 
distances a $U(N_c)$ four (three) dimensional gauge theory.
D6(5)-branes, when they intersect the 
D4(3)-branes, are interpreted as fundamental fields 
charged under the gauge groups. 
In \hw\ a particularly interesting phenomenon was 
noticed in which a D5-brane can pass through
an NS 5-brane and create a D3-brane.
This phenomenon has led to a discovery of a new phase transition between two
different gauge theories in $N=4$ supersymmetric gauge theory in three
dimensions. Concretely, the two theories are a $U(N_c)$ gauge theory coupled
to $N_f$ flavors in the fundamental representation and a $U(N_f-N_c)$ gauge 
theory 
coupled to $N_f$ flavors. 
When the two NS fivebranes adjoining the $N_c$ 3-branes interchange their
position, there is a transition from one theory to the other which passes
through the point of strong coupling for 
both theories. In this sense the two theories were
called in \hw\ a continuation of each other past infinite coupling.

Exploiting the methods of \hw, Elitzur, Giveon, and Kutasov \keg\ 
showed how one could move 
an NS 5-brane of type IIA string theory 
through D6-branes to continuously connect
a gauge theory of one rank, $SU(N_c)$, to 
a gauge theory of a different rank, $SU(N_f - N_c)$, 
manifesting Seiberg's duality in brane language.
In the 3d $N=4$ case, the theories have different Coulomb branches, and this
motion of the NS 5 branes is interpreted as a phase transition in which the
massless matter content changes during the transition.
In contrast, the 4d $N=1$ case has no Coulomb branch, and thus the Higgs branch
parameterizes all of moduli space. In this sense, this transition may be called
a duality. There are however problems with this interpretation which will be
discussed in the sequel.

\keg\ also showed how to add an adjoint chiral superfield, $X$,
to the brane picture and how to interpret
the deformations of a particular 
superpotential $W = \Tr X^{k+1}$
as the motion of $k$ NS 5-branes. 
They were able to connect
an $SU(N_c)$ gauge theory to an $SU(kN_f - N_c)$
gauge theory by continuously moving $k$
NS 5-branes past $N_f$ D6-branes finding a result 
that had been known from field theory \kut.

In section 2, of this paper, we will review the 
Elitzur, Giveon, and Kutasov construction and comment on 
the Higgs branch and the problem of realizing chiral symmetry.
In section 3, we will generalize the work of \hw\ and 
\keg\ by suspending two sets of D4-branes between three NS 5-branes,
thus constructing
product gauge groups $SU(N_c) \times SU(N_c')$ 
coupled with matter field $F$ in the $\bf {(N_c,\bar N_c')}$ 
representation and conjugate field $\tilde F$.
The theory in section 3 will have a superpotential
of the form $W = \Tr (F\tilde F)^2$.
By moving D4-branes, D6-branes, and NS 5-branes 
through the ten dimensional space of 
type IIA string theory and making \hw-type transitions, 
we will arrive at a new configuration of branes 
whose world volume theory is the dual described in \ils. 
After studying the simplest $N=1$ product duality in detail from 
the brane point of view, we will move on in section 4 
and consider the theory of 
\ils\ with a more general superpotential 
$W = \Tr (F\tilde F)^{k+1}$ 
and see how that translates into brane language.
In section 5, we
will add adjoint matter to 
both gauge groups and arrive at a duality of \brodie. 
Finally, in section 6, we will discuss the field theory and the 
brane configuration for a 
new duality involving three gauge groups and show how that 
naturally generalizes to a duality for $n$ gauge groups.

\newsec{Brane configurations}
\subsec{Supersymmetry.}
\subseclab{\secsusy}

The configurations we will study involve three
kinds of branes in type IIA string theory: 
a Neveu-Schwarz (NS) fivebrane,
Dirichlet (D) sixbrane and Dirichlet 
fourbrane.
Specifically, the branes are:

\item{(1)} NS fivebrane with worldvolume 
$(x^0, x^1, x^2, x^3, x^4, x^5)$, which lives at 
a point in the $(x^6, x^7, x^8, x^9)$ directions.
The NS fivebrane preserves supercharges of the 
form\foot{$Q_L$, $Q_R$ are the left and right moving 
supercharges of type IIA string 
theory in ten dimensions. They are (anti-) chiral: 
$\epsilon_R=-\Gamma^0\cdots\Gamma^9\epsilon_R$,
$\epsilon_L=\Gamma^0\cdots\Gamma^9\epsilon_L$.}
$\epsilon_LQ_L+\epsilon_RQ_R$, with 
\eqn\nsfive{
\eqalign{\epsilon_L=&\Gamma^0\cdots\Gamma^5\epsilon_L\cr
\epsilon_R=&\Gamma^0\cdots\Gamma^5\epsilon_R.\cr
}}

\item{(2)} D6-brane with worldvolume
$(x^0, x^1, x^2, x^3, x^7, x^8, x^9)$, which lives at a point
in the $(x^4, x^5, x^6)$ directions.
The D6-brane preserves supercharges satisfying
\eqn\dsix{
\epsilon_L=\Gamma^0\Gamma^1\Gamma^2\Gamma^3\Gamma^7\Gamma^8
\Gamma^9\epsilon_R.
}

\item{(3)}
D4-brane with worldvolume $(x^0, x^1, x^2, x^3, x^6)$
which preserves supercharges satisfying 
\eqn\dfour{
\epsilon_L=\Gamma^0\Gamma^1\Gamma^2\Gamma^3\Gamma^6\epsilon_R.
}

We will also use branes with different orientation in spacetime. One particular
brane will be called NS' brane and is rotated 90 degrees with respect
to the NS brane in item 1.
\item{(4)} \nsp\ fivebrane with worldvolume
$(x^0, x^1, x^2, x^3, x^8, x^9)$ preserving the
supercharges
\eqn\nsprime{
\eqalign{\epsilon_L=&\Gamma^0\Gamma^1\Gamma^2\Gamma^3
\Gamma^8\Gamma^9\epsilon_L\cr
\epsilon_R=&\Gamma^0\Gamma^1\Gamma^2\Gamma^3
\Gamma^8\Gamma^9\epsilon_R.\cr
}}

There are four supercharges
satisfying equations \nsfive-\nsprime;
${1\over 8}$ of the original supersymmetry of 
type IIA string theory is preserved. 
Each relation \nsfive-\nsprime\ by itself would break
${1\over 2}$ of the supersymmetry. Equations \nsfive\ and 
\dsix\ are independent and together break to ${1\over 4}$. 
Equation \dfour\ is not independent of \nsfive\ and 
\dsix\ and so breaks no more of the supersymmetry.
If we only had the NS 5-brane, the D6-brane, and the D4-brane
we could only make $N=2$ four-dimensional supersymmetric gauge theories as
were studied in \witten. However, by introducing the NS'-5 brane
as was done in \keg, we have a new equation 
\nsprime\ that is independent of \nsfive-\dfour. Altogether the branes preserve
${1\over 8}$ of the supercharges. In general we can consider
rotating the 5-branes to some arbitrary angle in 
$(x^4,x^5,x^8,x^9)$ \berkooz (see a related discussion in \barb).
As long as all 5-branes don't live exclusively in $(x^4,x^5)$, 
we will have $N=1$ supersymmetric configurations. 

Instead of an NS' 5-brane, we can rotate the 
6-branes to $(x^0,x^1,x^2,x^3,x^4,x^5,x^7)$, which will be called D' branes.
This gives us the relation
\eqn\dsixprime{
\epsilon_L=\Gamma^0\Gamma^1\Gamma^2\Gamma^3\Gamma^4\Gamma^5
\Gamma^7\epsilon_R.
}
The D'6-branes, together with the NS 5-branes and the D4-branes, break 
${1\over 8}$ of the
supersymmetry.
Of course we can have configurations with NS, NS' D, D' and D4 which will not
break the supersymmetry further.

The presence of these branes breaks space-time Lorentz group $SO(1,9)$ to
$SO(1,3)\times SO(2)\times SO(2)$, where the first group is the Lorentz group of
the 3+1 dimensional theory we want to study, the second factor is the rotation
group which acts on the $(x^4,x^5)$ coordinates, and the third factor is the 
rotation
group which acts on the $(x^8,x^9)$ coordinates.
The spinor representations of $SO(1,9)$ decompose as $[\bf 
{(2,1)+(1,2)}]_{\pm\pm}$.
For the {\bf 16} representation, out of these spinors, only one is not broken, 
and we
can choose it to be $\bf {(2,1)}_{++}$.
For the $\bf {\overline{16}}$ representation, the unbroken spinor is 
$\bf{(1,2)}_{--}$.
and thus both $SO(2)$
groups act on the spinors. As such, they are both global R-symmetries.
Their action on the spinors is identical, and thus we can find a linear
combination which does not act on the spinors.
This reduces the global symmetries into a $U(1)_R\times U(1)_g$ symmetry.
This symmetry is consistent with the global symmetry for the 
four dimensional $N=1$ supersymmetric theories that we are interested.
In fact, they are identified with the symmetries of the models.

\subsec{D-brane boundary conditions.}
\subseclab{\secbc}

As in \hw\ the four branes will be finite in the $x^6$ coordinate.
The effective low energy theory 
on the world volume of the D4-branes will thus be
macroscopically 3+1 dimensional. Because 4 supercharges are
preserved, we indeed see that the theories have $N=1$ supersymmetry in four
dimensions.
The finite D4-branes can end on either 5-branes or 6-branes, a fact which
can be deduced, by a sequence of T - and S - dualities from known
configurations.
The world volume of the D4 brane contains a $N=4$ $U(1)$ multiplet (with 16
supercharges).
When the D4-brane ends on the 5-brane or the 6-brane it
is subject to boundary conditions which project out 
some of the massless states from the theory.
This gives few cases in which different multiplets survive.
Below, we review, following \hw, what states exist for specific boundary
conditions on 4-branes. 

1) For a 4-brane between two NS 5-branes, there is a $N=1$ vector multiplet 
and 
chiral multiplet in the adjoint representation living 
on the world volume theory of the 4-brane (the adjoint chiral multiplet 
corresponds to motion of the 4-brane in 
the $(x^4,x^5)$ direction).
When $n$ D4-branes coincide, it seems that we get a 
$U(n)$ gauge theory with
a chiral adjoint matter. However, as in \witten, requiring finite energy
solutions on the 5-branes leads to freezing the $U(1)$ multiplet and instead
introducing a mass term for the adjoint field.
Thus we get a $SU(n)$ gauge group with chiral adjoint matter.

2) For a D4-brane between an 
NS 5-brane and an NS' 5-brane there is only a vector multiplet.
In general, the mass of the adjoint chiral multiplet that appeared in 
item one is proportional to the angle between the two 5-branes.
Since the NS 5-brane and the NS' 5-brane are at 
90 degrees, the mass of the adjoint is infinite.
For $n$ such D4-branes we get $SU(n)$ gauge theory.

3) For a D4-brane between two D6-branes, 
there are two chiral multiplets. These chiral multiplets are associated with
motion of the D4-brane along the $(x^7,x^8,x^9)$ coordinates 
together with the $A_6$
component of the gauge field. The $A_6$ component is compact with a radius
proportional to the gauge coupling.
These four scalars are conveniently written in
terms of two chiral multiplets $x_8+ix_9$; $x_7+iA_6$.

4) For a 4-brane between a D6-brane and NS' 5-brane
there is one chiral multiplet $x_8+ix_9$.
In general, for D6 brane and NS brane which are parallel, namely a rotated
configuration of the D6 NS' system, there is one chiral multiplet which
corresponds to motion of the D4 brane in between the two branes.

5) For a D4-brane between a D6-brane and an NS 5-brane
there are no massless moduli which contribute to the
low energy field theory. In general, this is true for 
any 5-brane that is not parallel to the 6-brane. 
For cases in which there are more than one brane in
between the 5-brane and 6-brane these states are called S-configurations.  

In addition to the above massless states, we will need also other types of
massless states coming from configurations in which D6-branes meet D4-branes in
space.
Fundamental type IIA strings stretching between the 4-branes and the 6-branes
look like particles to an observer living on the 4-brane.
The endpoints of the strings are electrically charged with respect to the
gauge group on the D4-brane.
These strings are the 
fundamental fields $Q$ and $\tilde Q$, the quarks. 

In the configuration in item five, in which a NS brane is not parallel to a D6
brane, the branes meet in space whenever their $x^6$ positions coincide. That is
they meet on a line. If they pass 
through each other a D4 brane is created as in
\hw. This is a slight generalization of the configuration discussed in \hw\ to
an arbitrary angle, excluding one case in which the branes are parallel.
In the case that the branes are parallel, they meet in space whenever three of
the coordinates are tuned. In this case the branes can easily avoid each other
when changing the $x^6$ coordinate by passing through the other coordinates.
Thus a D4 brane will not be created during such a transition.
In this sense the case when the branes are parallel is a special case.
We may ask, however, what happens when the branes do meet is space by tuning
their three coordinates to coincide. We will argue in section 2.4 that this
leads to enhancement of gauge symmetry on the world volume mutual to both the
branes which is interpreted as enhanced global chiral symmetry on the world
volume of the D4 brane.

This discussion suggest two cases. Either the NS brane and D6 brane are parallel
and give rise to enhanced chiral symmetry when they meet in space or the branes
are not parallel and when they pass through each other a D4 brane is created.

\subsec{Review of Elitzur, Giveon, and Kutasov's transition}
\subseclab{\secKEG}

\ifig\figseiberg{Vertical solid lines point in the $(x^4,x^5)$ direction. 
Vertical dashed lines point in the $(x^8,x^9)$ direction. 
The horizontal direction is $x^6$. The electric theory is shown on 
the left, the magnetic theory on the right.}
{\epsfxsize4.0in\epsfbox{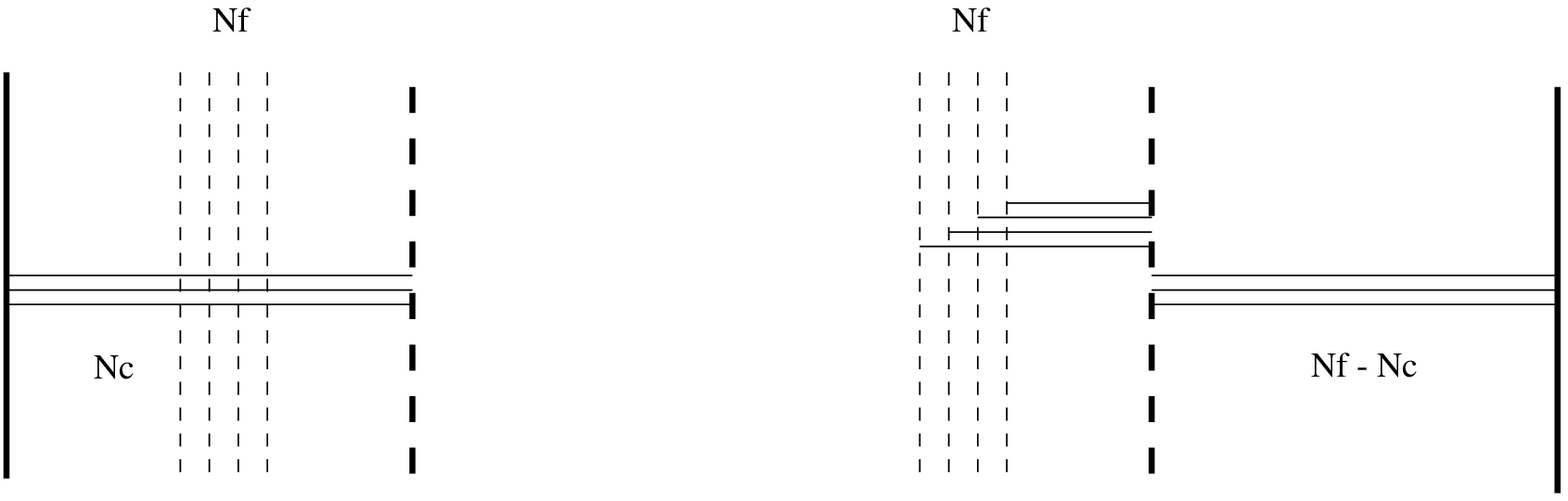}}

First let's review what was done in \keg. The brane
configuration of \keg\ is
shown in \figseiberg. In the electric configuration shown on the left, 
we have $N_c$ D4-branes stretched between 
an NS 5-branes on the left and an NS' 5-brane on the right. 
In addition there are $N_f$ D6-branes with an $x^6$ coordinate in between the
two coordinates of the 5-branes.
According to our rules above, this gives us $SU(N_c)$ gauge theories with $N_f$
fundamental flavors in four-dimensions

\figseiberg\ 
shows D4-branes which are frozen in space; there are no moduli
associated with this configuration. This is consistent with the fact that $N=1$
supersymmetric gauge theories have no Coulomb branch.
The Higgs branch was described in \hw\ for twice the amount of supersymmetry,
and we can repeat the analysis here. Transition to the Higgs branch is done by
moving the D6-branes in the $(x^4,x^5)$ coordinates to a
point where they touch the D4-branes. The quarks become massless and a
transition to the Higgs branch is possible by breaking the D4-branes along the
D6-branes and moving in the $(x^8,x^9)$. 
\ifig\fighiggs{The Higgs branch of $N=1$ supersymmetric QCD versus the Higgs
branch of $N=2$ supersymmetric QCD. The numbers which are assigned to each
D4-brane denote the number of chiral superfields which are associated with
motion of the D4-brane along the branes at its ends.
In the bottom picture, the $N=2$ case, some of the branes, which carry no
moduli, are not broken to avoid S-configurations.
In the top picture, such a restriction does not exist for the NS' brane.
Instead there are extra moduli as indicated in the figure.
}
{\epsfxsize4.0in\epsfbox{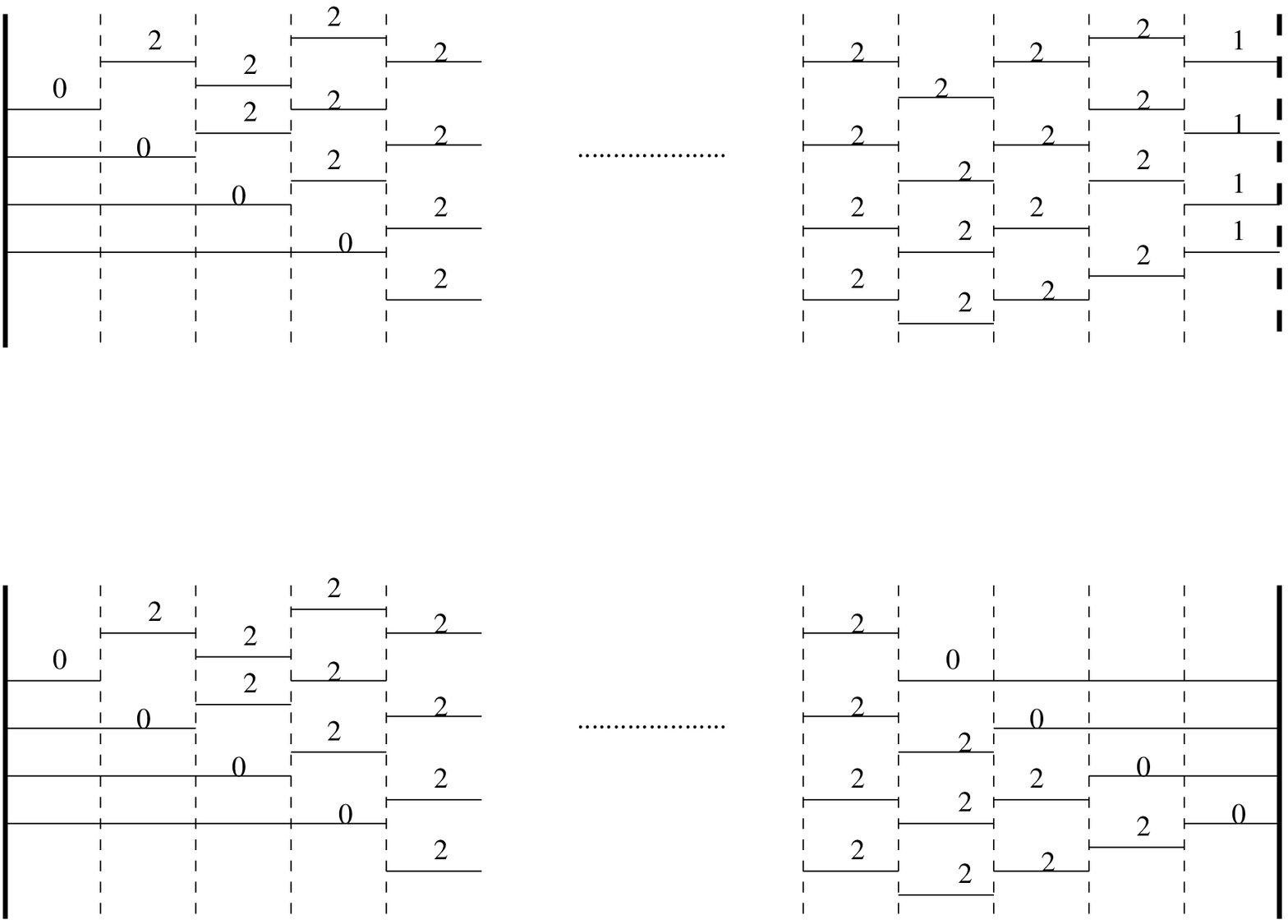}}
\fighiggs\ shows the maximal Higgsing possible. 
If one counts the number of chiral multiplets according to the 
rules above, we see that the dimension of the Higgs branch is 
$2N_fN_c - N_c^2$ for $N_f > N_c$ and $N_f^2$ for $N_f \leq N_c$ 
in agreement with field theory.

Now we consider the transition caused by exchanging the two NS branes.
We first move the NS 5-brane past the
D6-branes where the transition described in \hw\ occurs.
$N_f$ 4-branes are created stretching between the $N_f$ 6-branes on 
the left and the NS 5-brane on the right. 
As in \hw\ in order to avoid magnetic charge transition between the
two 5-branes, we move the NS in the $x^7$ direction.
This has the effect of
turning on a FI term and moving to the Higgs branch.
If there are D4-branes stretched between the two NS branes, 
such a process
changes the orientation of the D4-branes in space and thus 
breaks supersymmetry.
In order to avoid such a thing 
we need to reconnect $N_c$ of the 4-branes, which
are stretched between the 6-branes and the NS brane, to all of the 4-branes
which are between the NS brane and the NS' brane. Such a 
reconnection corresponds
to moving the 4-branes in the $(x^4,x^5)$ 
direction, and thus to massless quarks.
At this point there are no 4-branes between the NS brane and the NS' 5-brane, 
and thus it can
be moved in the $x^7$ direction without breaking the supersymmetry. 
Then we can exchange the position of the NS branes in the $x^6$ direction and
move back to the origin of the Higgs branch by setting the $x^7$
coordinate of the NS brane to zero.
Note that in this process the NS' brane plays a passive role, namely it stays
in place while all other branes move around. As such this transition can be
connected to the transition in \hw\ by a rotation of the NS brane to NS' brane
after T-duality in the 3rd coordinate.

We end up with the configuration on the right in \figseiberg. 
 From left to right in $x^6$, 
$N_f$ 4-branes stretch between the $N_f$ 6-branes and the NS' 
5-brane, and $N_f - N_c$ 4-branes stretch between the NS' and the NS.
Here we have Seiberg's dual $SU(N_f-N_c)$ gauge group. The strings 
stretching between the two sets of 4-branes are the dual quarks $q$ and
$\tilde q$. The 4-branes can break completely along the rightmost 6-brane,
 and thus
generate $N_f$ 4-branes which move freely in the 89 directions. These branes
give rise to $N_f^2$ singlet fields $M$ which are the states stretched
between the $N_f$ 4-branes in the $(x^8,x^9)$ direction. Only $N_f$ fields,
which correspond to the ``Cartan generators" of these mesons are visible in the
brane picture.
Moving D6-branes in the 
$(x^4,x^5)$ direction corresponds to giving a
flavor a mass in the electric theory while this motion corresponds 
to Higgsing in the magnetic theory. Breaking a D4-brane on a D6-brane and
moving the D4-brane in the $(x^8,x^9)$ direction corresponds in the 
electric theory to 
Higgsing. Motion of one of the $N_f$ D4-branes in the $(x^8,x^9)$ 
direction in the magnetic 
theory corresponds to giving a fundamental $Q$ field a mass. In this way, we 
can see that there is a superpotential $W = Mq\tilde q$.

\ifig\fighiggsdual{The Higgs branch of Seiberg's dual $SU(N_f - N_c)$
gauge theory. The dimension of the dual Higgs branch is 
the same as the dimension of original Higgs branch $2N_fN_c - N_c^2$.}
{\epsfxsize4.0in\epsfbox{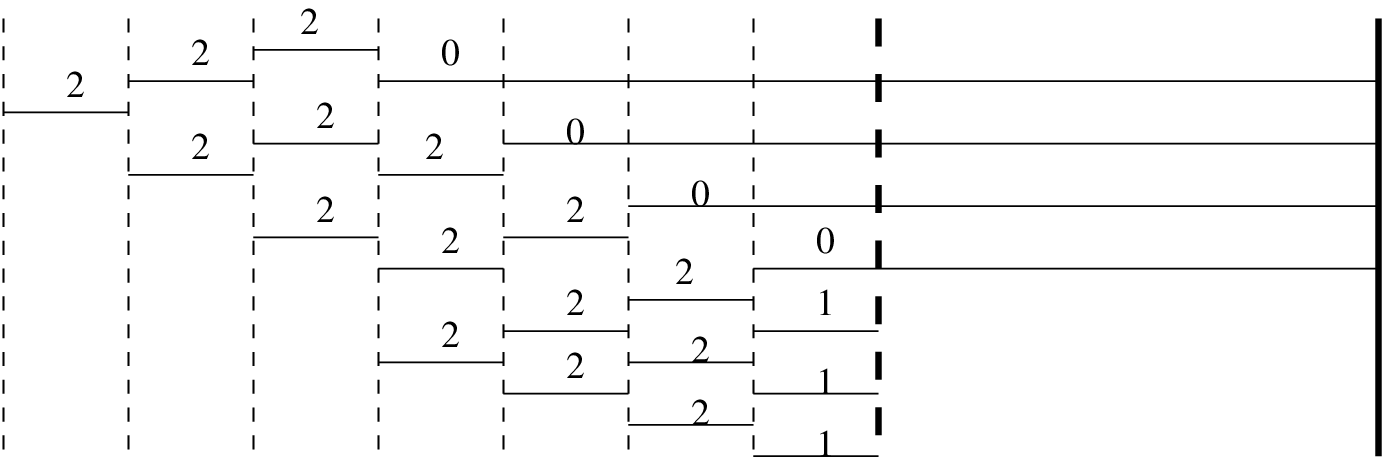}}

It is interesting to check that the dual Higgs branch shown in \fighiggs\
has the same dimension as the original Higgs branch shown in \fighiggsdual.
This is a necessary condition if these two theories are to be 
called dual.

There are a few comments in order.
In making the transition from one theory to another, the gauge coupling changes
continuously and passes through infinite coupling of both theories. In this
sense this process is a phase transition not a duality. One the other hand,
the space of vacua of these theories is parameterized by Higgs
expectation values which give, as is clear from the brane construction,
identical moduli spaces in both theories. In this
sense the name duality is justified.

Another problem is with the interpretation of the $x^7$ coordinate as a 
FI term for the $U(1)$ gauge group.
In \witten\ it was argued that the gauge group which naively
looks like $U(N_c)$ gauge theory is really $SU(N_c)$ by requiring finite energy
solutions of the 5-branes. This restriction removes the $U(1)$ from the theory
and thus does not allow for a FI term to appear as a parameter in the field
theory, according to
conventional wisdom. From the point of view of the branes, this is nevertheless
a well defined procedure which by analogy to the $d=3$ $N=4$ case bares the name
FI term.

Another way to calculate the dual gauge group is to use linking numbers. 
As in \hw, we assign a number to each NS 5-brane and D6-brane depending on 
the types of branes to its left or right. 6-branes or 5-branes to the 
right of the
brane under consideration contribute $-1\over 2$ linking number while 
6-branes or 5-branes to the left contribute $+1\over 2$ linking number.
4-branes to the right contribute $+1$ linking number while
4-branes to the left contribute $-1$ linking number.
These linking numbers were assigned in the $N=4$ $d=3$ setup. However by
T-duality, we can assign these numbers to the $N=2$ $d=4$ setup.
Now the motion of branes can be calculated for the 8 supercharges case, namely
when the NS branes are parallel. Once the transition is done it is possible to
make a 90 degree rotation of the NS brane to become an NS' brane.
This process corresponds to adding a mass term to the adjoint and thus lead
from the $N=2$ system to the $N=1$ system in the spirit of \asp\
(for a related discussion, see \barb).
The rotation of the NS brane, which is done without encountering any other brane
in this process preserves the magnetic charge of the brane, or the 
linking number,
and thus allows us to use these linking numbers as conserved quantities of these
transitions.
In the brane configuration described above, the NS 5-brane started off on the 
left with linking number $-{N_f\over 2} + N_c$. After it moved over to 
the right, it's linking number became $+{N_f\over 2} - \tilde N_c$. Since
the linking number for all branes must be conserved, the dual 
gauge group is $\tilde N_c = N_f - N_c$. 

In \keg, they also discussed a brane configuration for another
field theory duality proposed by Kutasov and Schwimmer 
in \kut. This theory is an $SU(N_c)$ gauge theory with 
$N_f$ fundamentals $Q$ and $\tilde Q$, an adjoint field $X$, and a 
superpotential $W = \Tr X^{k+1}$. The brane configuration is the
same as above except we have $k$ NS 5-branes instead of a single NS 5-brane.
Relative motion of the NS 5-branes in the $(x^8,x^9)$ directions 
corresponds to turning on lower order operator
in the superpotential, giving the adjoint field a vacuum expectation value, 
and breaking the gauge group to $k$ copies of Seiberg's duality. 

\subsec{Realization of Chiral Symmetry}
\subseclab{\secchiralsym}

A problem with the brane construction of \keg\ is that it does not 
have the same global symmetries as the theory of Seiberg, namely, it 
lacks chiral symmetry. Because of this, the brane configuration of 
Elitzur, Giveon, and Kutasov resembles a theory that ``remembers" it's 
$N=2$ origin. In this section, we discuss how this problem with 
the brane configuration can be corrected.

If we look at the $N=2$ supersymmetric QCD, we notice that there is 
a term in the superpotential, 
\eqn\wneqtwo{W = \lambda QX\tilde Q.} $\lambda$ is fixed to $\sqrt 2$ by the
supersymmetry.
This term requires that when we give a vacuum expectation value to the 
adjoint field, the hypermultiplets receive a mass.
This term also breaks the global chiral symmetry from 
$SU(N_f)_R \times SU(N_f)_L$
to $SU(N_f)$.
Turning off the coefficient $\lambda$ in \wneqtwo, 
breaks the $N=2$ supersymmetry to $N=1$ while restoring
the chiral symmetry.
Chiral symmetry can be broken explicitly whenever the quarks receive a mass
or dynamically when the quarks receive a vacuum expectation value.
How can we see this chiral symmetry in the brane picture?
$N=2$ supersymmetric QCD corresponds in brane language to 
two parallel 5-branes extended in the $(x^4,x^5)$ direction and 
$N_f$ 6-branes extended in the $(x^8,x^9)$ direction. $N_c$ 4-branes 
connect the 5-branes. Motion of the 4-branes in the $(x^4,x^5)$ direction 
corresponds to giving the adjoint field a vacuum expectation value 
and Higgsing the 
gauge group to it's Cartan subalgebra. Moving the 6-branes in the 
$(x^4,x^5)$ 
direction corresponds to giving the hypermultiplets a mass, so we naturally
see that the adjoint field is coupled to the hypermultiplets 
as in \wneqtwo.
If we consider rotating the 6-branes from extending in the $(x^8,x^9)$
to extending in the $(x^4,x^5)$ direction, we see that now giving a mass
to the hypermultiplets corresponds to motion of the 6-branes in the 
$(x^8,x^9)$ directions. In the brane picture, the adjoint and 
the fundamentals have decoupled in the superpotential. 
This rotation of the 6-branes 
corresponds \ah\ to turning off the parameter $\lambda$ in equation \wneqtwo.
Now we should be in a position to see the chiral symmetry. We propose that 
chiral symmetry is restored when the 6-branes touch a parallel 5-brane. 
The only direction that the 6-branes and the 5-brane do not share is $x^7$.
Therefore, a 6-brane can have a boundary on the 5-brane in the $x^7$ direction.
Indeed, this was proposed in \branesendingI,\branesendingII.

To support this proposal, we note that the distance between, say a NS' brane and
a D6 brane is a 3-vector in $(x^4,x^5,x^6)$ space. To make the two branes 
coincide, we need
to tune the three coordinates in these directions. When they touch, a massless
multiplet comes down in the world volume $(x^0,x^1,x^2,x^3,x^8,x^9)$ 
which is mutual to both branes.
This world volume is six dimensional, 
and thus we expect to get a massless state
in six dimensions when both branes meet in space.
We also note that locally, in the absence of D4 branes, the supersymmetry is
broken to 8 supercharges, 
and thus we are dealing with $(0,1)$ supersymmetry in
six dimensions.
There are two possible types of multiplets which may become massless at the
point when the branes meet: a hypermultiplet and a vector multiplet.
Supersymmetry implies that vector multiplets become massless when three FI
parameters are set to zero. These are the three distances mentioned above, and
thus this rules out the possibility of having a massless hypermultiplet at the
point in question.
The distance in the $(x^4,x^5,x^6)$ directions now gets the interpretation of a 
FI
parameter for a $U(1)$ gauge field that becomes massless when the two branes 
meet
in space.

In the presence of D4 branes the local supersymmetry is broken from 8
supercharges to 4 supercharges, as well as the local Lorentz invariance from
$SO(1,5)$ to $SO(1,3)$. The $U(1)\times U(1)$ gauge symmetry on the six
dimensional world-volume theory becomes now global $U(1)\times U(1)$
chiral symmetry of the four dimensional theory.

This setup can be generalized in the following way. When $N_f$
6-branes touch a 5-brane, they split in half, 
and the $(x^0,x^1,x^2,x^3,x^8,x^9)$ 
world volume theory becomes
$SU(N_f)_R\times SU(N_f)_L$. Strings stretching from the 
$N_c$ 4-branes to the $N_f$ 6-branes to the right in $x^7$, 
are the quarks $Q_R$
charged under $(\bf {N_c,N_f,1})$. Strings stretching from the 
$N_c$ 4-branes to the $N_f$ 6-branes to the left in $x^7$, 
are the quarks $Q_L$
charged under $(\bf {\bar N_c,1,N_f})$. 
Moving the 6-branes off the 5-brane in the 
$(x^4,x^5)$ direction breaks the six dimensional symmetry spontaneously while
breaking the four dimensional chiral symmetry explicitly, as it should 
since motion in the $(x^4,x^5)$ corresponds in the field theory 
to giving the quarks a mass. Breaking the 4-branes on the 6-branes
is only possible if the 6-branes move off the 5-brane in the 
$(x^8,x^9)$ directions. This corresponds to  
Higgsing, and in agreement with field theory, Higgsing breaks 
chiral symmetry.

The string theory setup also predicts a phenomenon which is hard to understand
from a field theory point of view. This corresponds to moving the 6-branes in
the $x^6$ direction. This motion, like the $(x^4,x^5)$  
motion, corresponds to breaking
the six dimensional symmetry spontaneously while breaking 
explicitly the four
dimensional chiral symmetry. The parameter which governs that is the $x^6$
distance between the NS' brane and the D6 brane which is real. Such a parameter
is problematic since it has no natural complex partner and thus violates
holomorphy if it appears in the superpotential. Thus we expect it to appear
only in D-terms. On the other hand such a parameter has no natural
interpretation in the field theory. A similar problem was encountered in \hw\
where ``magnetic couplings" where introduced by mirror symmetry which did not
have a natural interpretation in the field theory setup. In \hw\ it was also
suggested that the relative distance between NS and D branes is irrelevant for
the field theory. This is also valid when the branes are oriented in any angle
which is not the parallel case.
In this case we see that, when the NS and D branes are
parallel, these parameters become relevant as they control the breaking of the 
chiral symmetry of the theory.

Finally for this section, one may wonder what are the objects for which
quantization leads to the appearance of massless vector multiplets.
As is, by now well known, quantization of open strings lead to massless
hypermultiplets whenever two D-branes (which break to 1/4 of the supersymmetry)
meet in space. It is not known however what are the states which get massless
when a D brane meets a NS brane. The above analysis predicts that the states
are vector multiplets. Now the only virtual states which end on both a NS brane
and a D6 brane are D4 branes. There is no other brane which has this property,
on the other hand, we have been using this property for D4 throughout 
this paper.
Thus we can have virtual open D4-branes which have threebrane boundaries which
propagate on the worldvolume of the D6 and NS branes.
To a D6 observer they look like monopoles while for a NS observer they look like
vortices. When these two branes touch, the tension of the threebranes vanishes.
The worldvolume of the D4-branes consists of 0123 and a real line in the 456
space which connects the NS and D6 branes.
A supersymmetric configuration which is consistent with the supersymmetries in
this problem implies that the D6 and NS branes will have identical 45 positions
and different $x^6$ positions. This again gives a special case to the point
where in the field theory the masses are zero and chiral symmetry is expected.
Thus we are led to predict that quantization of tensionless threebranes in six
dimensions gives rise to massless vector multiplet in six dimensions!
It would be very interesting to provide further support for this scenario.

\newsec{Duality with two gauge groups: $SU(N_c)\times SU(N_c')$}
\subsec{The theory with superpotential $W = \Tr (F\tilde F)^2$.}
\subseclab{\sectwogroups}

\ifig\figils{
This figure is in $(x^4,x^8,x^6)$ space.
The thick lines are the 5-branes which point at arbitrary angles 
in $(x^4,x^8)$ while the 4-branes marked by their numbers, $N_c$ and $N_c'$, 
point in the $x^6$ direction.}
{\epsfxsize2.0in\epsfbox{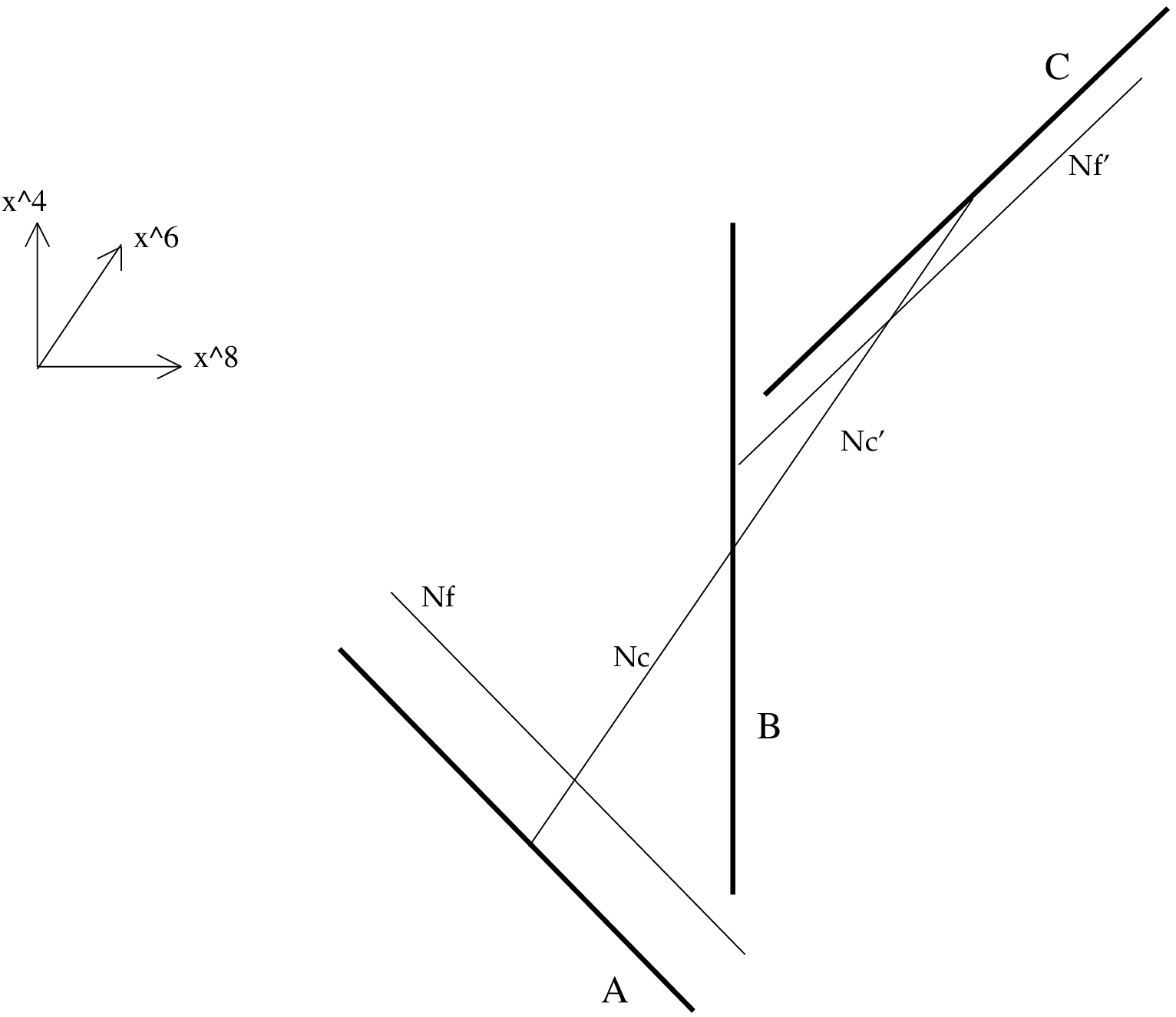}}

In this section,
unlike in section \secKEG, 
we will find we cannot suffice with only
5-branes oriented at 0 degrees $(x^4,x^5)$ and 90 degrees
$(x^8,x^9)$. We will need 5-branes at arbitrary angles 
in $(x^4,x^5,x^8,x^9)$. 
Therefore, referring to the 5-branes as NS and NS' 
no longer makes sense. In this section we will label 
the 5-branes with letters $A$, $B$, and $C$. The $B$ 5-brane is in the 
middle of the other two 5-branes and is oriented
at zero degrees (i.e. it points along the $(x^4,x^5)$ direction).
The $A$ 5-brane is oriented at angle $\theta_1$ with respect to 
the $B$ 5-brane, and the $C$ 5-brane is oriented at 
an angle $\theta_2$ with respect to the middle $B$ 5-brane.
We orient the $N_f$ 6-branes in the direction parallel to
the $A$ 5-brane, and the $N_f'$ 6-branes in the direction parallel to
the $C$ 5-brane. In this way we get the brane realization of chiral symmetries
which are present in the field theory. The configuration is shown in \figils.

\ifig\figtwo{This figure is a different view of \figils.
The vertical lines point in $(x^4,x^5,x^8,x^9)$.
The horizontal lines point in $x^6$.
$SU(N_c)\times SU(N_c')$
gauge fields live on the D4-branes. Strings stretching between the 
two sets of 4-branes are the fields $F$ in the $\bf {(N_c, \bar N_c')}$ 
representation. There are strings stretching from $N_c$ 4-branes to the 
$N_f$ 6-branes. These are the fundamental fields $Q$ and $Q'$. There
is a superpotential $W = \Tr (F \tilde F)^2$.
The strings drawn near B are somewhat misleading since these 
strings are really inside the NS brane. It must be so because of BPS condition 
and minimal length criterion.}
{\epsfxsize2.0in\epsfbox{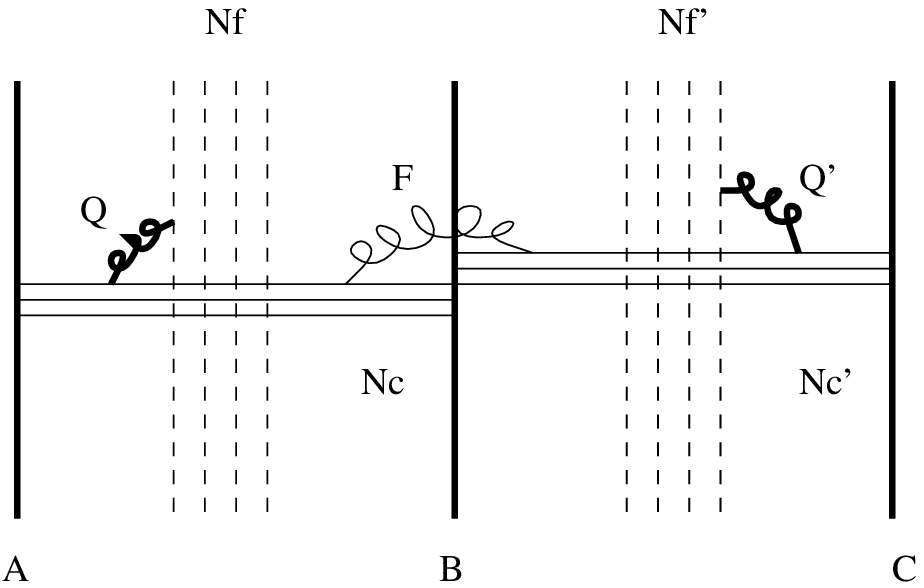}}

Let's consider \figtwo. 
We have 
from left to right, $N_c$ 4-branes stretched between the $A$ 5-brane and the 
$B$ 5-brane which is connected to the $C$ 5-brane by $N_c'$ 4-branes.
This gives us an $SU(N_c)\times SU(N_c')$ gauge theory.
$N_f$ 6-branes intersect the $N_c$ 4-branes, 
and $N_f'$ 6-branes intersect the $N_c'$ 4-branes in the $x^6$ direction.
Strings stretching between the $N_f$ 6-branes and the $N_c$ 4-branes 
are chiral multiplets $Q$ and $\tilde Q$ in the fundamental representation
of $SU(N_c)$, while strings stretching between the $N_f'$ 6-branes and the 
$N_c'$ 4-branes 
are the chiral multiplets $Q'$ and $\tilde Q'$ in the fundamental 
representation of $SU(N_c')$. Strings can also stretch between the
$N_c$ 4-branes and the $N_c'$ 4-branes. These fields we will call
$F$ and $\tilde F$ and are in the $\bf {(N_c,\bar N_c')}$ and  
$(\bf {\bar N_c, N_c')}$ representation of the gauge group respectively.
We will argue below that this theory has a superpotential
$W = \Tr (F \tilde F)^2$.
The field theory that this brane configuration corresponds to 
was described in \ils. There it was shown 
that the dual gauge group is $SU(2N_f'+N_f - N_c')
\times SU(2N_f + N_f' - N_c)$ with singlet fields that are a one-to-one 
map of the mesons of the electric theory. Let's see how this works in the 
brane language.  

\ifig\figtwodual{
This theory is the dual of \figils. It has $SU(2N_f' + N_f - N_c')
\times SU(2N_f + N_f' - N_c)$
gauge fields living on the D4-branes. There are hypermultiplets $F,\tilde F$
in the $\bf {(\tilde N_c, \bar {\tilde N_c'})}$ representation coming from 
strings
stretched along the B NS brane.
There are $N_f, (N_f')$ strings stretching from a fourbrane to the 
$\tilde N_c' (\tilde N_c)$ 4-branes along the A (C) 5-brane.
These are the fundamental fields $q', \tilde q' (q, \tilde q)$. The
gauge theory also has singlet fields coming from 
4-branes stretched between the NS' 5-branes and the 6-branes.}
{\epsfxsize3.0in\epsfbox{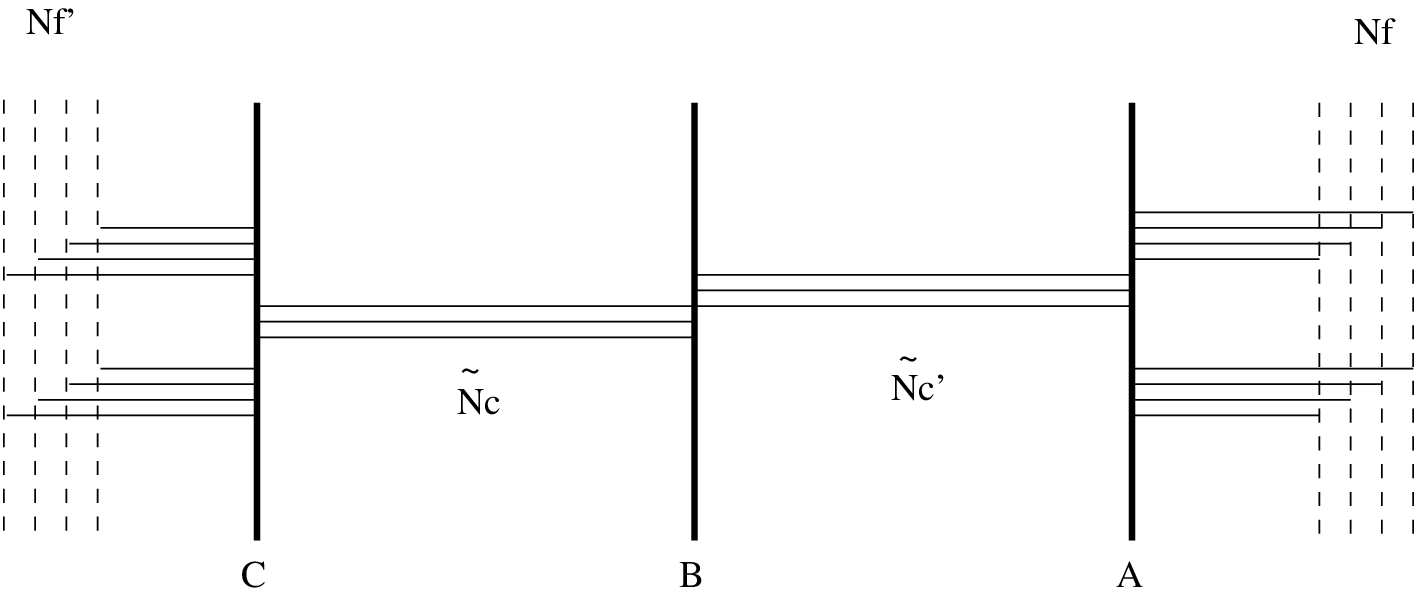}}

One way to find the dual theory is to use linking numbers as 
we did for the Seiberg duality in section \secKEG. 
Although in \hw\ the
linking numbers were only used for NS 5-branes intersecting 
6-branes at 90 degrees,
we claim that a 4-brane is created whenever a 5-brane crosses 
a 6-brane as long as the 5-brane and the 6-brane are not parallel.
The $N_f$ D6-branes start out with 
linking number ${-1\over 2}$. When we move them to the right past the 
$B$ and $C$ 
5-branes, their linking number becomes ${+3\over 2}$. Since the linking number
for a particular brane must be conserved, we must add two 4-branes to the 
left side of all $N_f$ 6-branes. These are the singlets corresponding
to the mesons $Q\tilde Q$ and $Q\tilde F F\tilde Q$. The singlets 
corresponding to the mesons $Q\tilde F Q'$ and $\tilde Q F \tilde Q'$ 
presumably become visible in the IR when the $N_f$ and $N_f'$ 6-branes 
intersect.
The $A$ 5-branes on 
the far left starts out with linking number 
$L = -{N_f\over 2} - {N_f'\over 2} + N_c$. 
After the duality, the $A$ 5-brane
ends up on the far right with linking number 
$L = -{N_f\over 2} + {N_f'\over 2} 
- \tilde N_c' + 2N_f$. In order to conserve linking number the dual
gauge group must be $\tilde N_c' = 2N_f + N_f' - N_c$. We can consider 
what happens when we move the $C$ 5-brane from the right to the left.
This gives us the other dual gauge group $\tilde N_c = 2N_f' + N_f - N_c'$. 
It is easy to check that the linking numbers for the middle $B$ 5-brane 
are consistent with this dual configuration.
The dual brane configuration is shown in \figtwodual.

A point on the relation between chiral symmetry and the singlet mesons is now
in order. In item 5 of section 2.2 we stated that there are no moduli for branes
which are not parallel. On the other hand in item 4 of that section there is a
chiral multiplet whenever the branes are parallel. These states are precisely
the states which give the chiral mesons of the dual theory as can be seen by
breaking the D4 branes to the right of the $A$ NS brane of figure 6.
Had we chosen the $N_f$ D6 branes to be nonparallel to the A NS brane, such a
breaking would lead to S-configuration which is assumed to break supersymmetry.
On the other hand for the parallel case, we have chiral mesons and enhanced
chiral symmetry when the branes meet in space. This is in perfect agreement with
the field theory expectations which thus provides further support to our claim
for enhanced chiral symmetry.

Another (more tedious) way to see the dual configuration is to 
dualize each gauge group independently of the other. We can label
each 5-brane from left to right $A,B,C$. We then move $B$ past $A$ which 
constitutes a Seiberg duality on the first gauge group. 
We have the configuration $BAC$. 
Then we perform another Seiberg duality on the second gauge group
which brings us to $BCA$. We dualize again the first (left most) 
gauge group and end up with $CBA$. We have effectively switched
the two end 5-branes while keeping the middle 5-brane in place. 
It is not hard to see that 
in the process, we have switched the D6-branes and put them outside of the  
5-branes. It is curious to note that although there are many ways 
we can arrange the 6-branes and 
5-branes, the dual turns out to be the one which is the ``mirror'' of the 
original configuration. We will see that this is true for theories with
arbitrary numbers of gauge groups.

\subsec{Derivation from $N=2$}
\subseclab{\secneqtwo}

The brane configuration of section \sectwogroups\ can be derived 
from an $N=2$ configuration. In field theory language, we begin with an 
$SU(N_c)\times SU(N_c')$ gauge theory with $N_f$ hypermultiplets charged under
$SU(N_c)$, $N_f'$ hypermultiplets charged under $SU(N_c')$, and a 
hypermultiplet in the $(\bf {N_c,N_c'})$ representation. 
$N=2$ supersymmetry forces us to have a superpotential
of the form 
\eqn\neqtwo{W = \lambda_1 QX_1\tilde Q + \lambda_2 Q'X_2\tilde Q' 
+ FX_1\tilde F + FX_2\tilde F.}
Breaking the $N=2$ supersymmetry, we set $\lambda_1 = \lambda_2 = 0$
rather than to $\sqrt{2}$.
Giving the adjoint fields $X_1$ and $X_2$ masses, $m_1$ and $m_2$,
respectively and integrating them out, we are left with
\eqn\neqone{W = 
-{1\over 2}\left( {1\over m_1} + {1\over m_2}\right) \Tr (F\tilde F)^2.}
Notice that if $m_1 = -m_2$ then the superpotential vanishes and 
if we take
the limit that the masses are infinite, then the superpotential again 
vanishes. 

\ifig\figtrig{We see the NS 5-brane configuration in the $(x^4,x^8)$
plane. The brane configuration on the right has a superpotential
$W = (F\tilde F)^2$. The configuration on the left
is the same theory perturbed by a mass term
$ W = (F \tilde F)^2 + \mu F \tilde F$.  
The dotted lines are not branes.}
{\epsfxsize3.0in\epsfbox{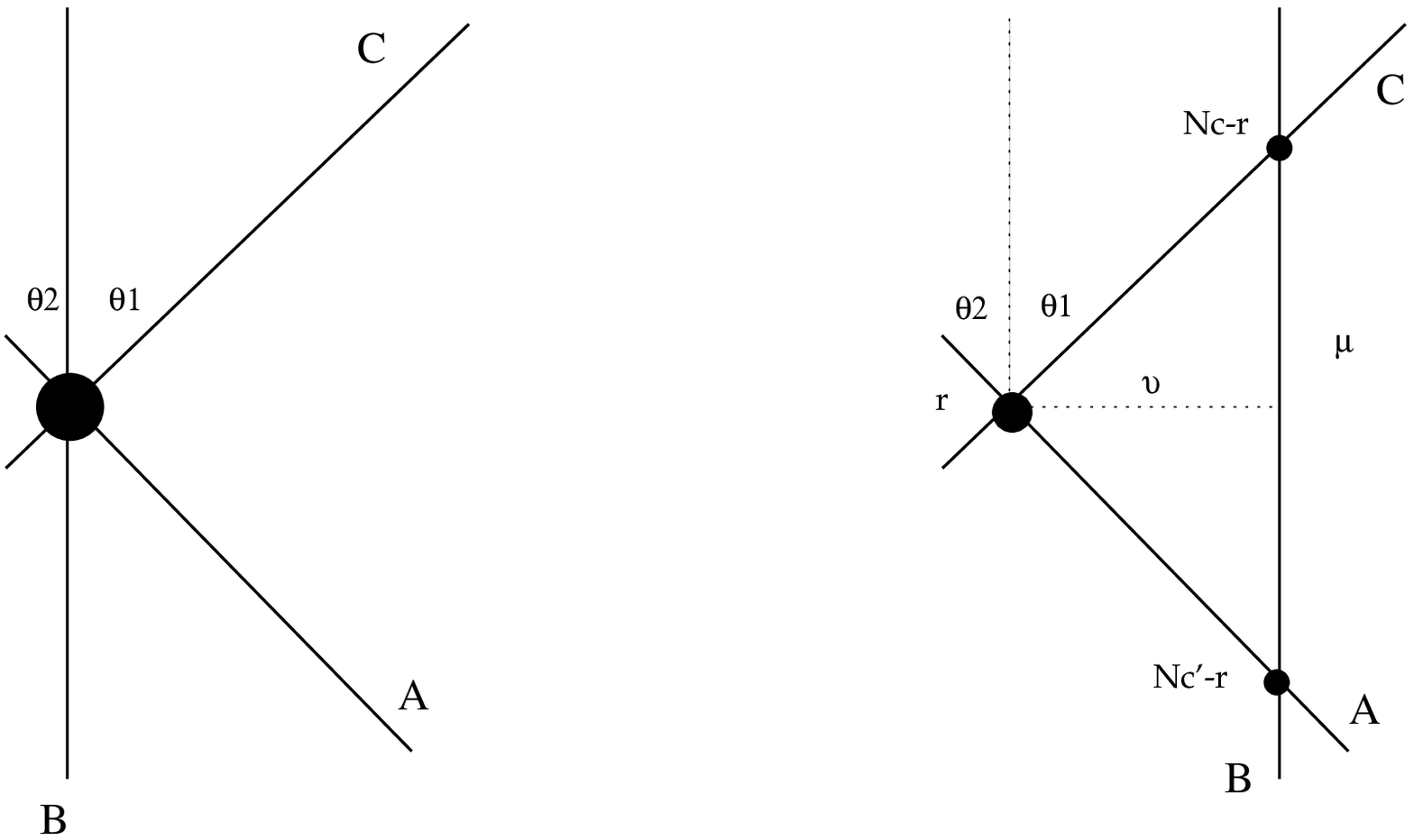}}

In brane language, the derivation of the superpotential amounts to beginning 
with the brane configuration described in section \sectwogroups;
the only difference being, as explained 
in section \secsusy, for $N=2$ supersymmetry 
we must have all 5-branes parallel in the $(x^4,x^5)$ 
with the 6-branes extending 
in the $(x^8,x^9)$ directions. 
We can rotate the 6-branes to 6'-branes 
such that they extend in the 
$(x^4,x^5)$ directions. This corresponds 
in the field theory to setting the terms $\lambda_1$ and 
$\lambda_2$ to zero as explained in \secchiralsym. We now rotate the NS 
5-branes. 
The mass of the adjoint field on the 
4-brane is proportional to the angle between the adjoining 
NS 5-branes. We do not want to rotate 
the 5-branes 90 degrees since this would give the 
adjoint fields infinite mass and drive the resulting 
superpotential term \neqone\ to zero. We also don't want to 
have the angles between the 5-branes 
equal and opposite since that also 
would result in $m_1 = -m_2$ and \neqone\ equaling zero.
We
rotate each end 5-brane and all 6'-branes an angle $\theta_1$ 
and $\theta_2$ with respect to the center 5-brane such 
that 
$0<\theta_1 <{\pi \over 2}$ and  
$0<\theta_2 <{\pi \over 2}$ as well as
$\theta_1 \neq -\theta_2$.
The final configuration is shown in \figils\ and \figtrig.

\subsec{Giving the superpotential a mass}
\subseclab{\secmass}

As was done in \ils, we can consider what happens when we give the 
superpotential a mass 
\eqn\mass{W = -{1\over 2}({1\over m_1} + {1\over m_2})
\Tr (F\tilde F)^2 + \mu \Tr F\tilde F.} 
The equations of motion 
for equation \mass\ are $F\tilde FF - \nu F = 0$ and 
$\tilde FF\tilde F - \nu \tilde F = 0$
where we have defined $\nu = {\mu\over {1\over m_1} + {1\over m_2}}$.  
A general solution to the supersymmetric vacuum condition is 
to have $<F\tilde F> = \nu$; 
breaking the gauge group $SU(N_c)\times SU(N_c')$
to $SU(N_c-r)$ with $N_f$ flavors, $SU(N_c' -r)$ with $N_f'$ flavors,
and $U(r)$ with $N_f+N_f'$. 
What happens in 
the brane configuration? 
The center $B$ 5-brane moves in the $(x^4,x^5)$ direction such that 
instead of the three 5-branes intersecting at a point in 
$(x^4,x^5,x^8,x^9)$ they intersect in three points. $N_c-r$ 4-branes move up 
the sloping $C$ 5-brane and connect to the middle $B$ 5-brane and 
intersect $N_f$ 6-branes. $N_c'-r$ 
4-branes move down the sloping $A$ 5-brane and intersect
$N_f'$ 6-branes. $r$ 4-branes remain in place 
and connect from the $A$ 5-brane
to the $C$ 5-brane intersecting $N_f + N_f'$ 6-branes.
This agrees nicely with what happens in field theory. 
We can even check the trigonometry! The distance from the $r$ 4-branes
to the middle $B$ 5-brane is $\nu$. The distance from the $N_c - r$ 4-branes
to the $N_c' - r$ 4-branes is $\mu$. Using elementary trigonometry 
and \figtrig\ it is
not difficult to show that 
\eqn\trig{\mu = \nu({1\over tan(\theta_1)} + {1\over tan(\theta_2)}).}
 From the relation for the adjoint fields mass, $m_1 = tan(\theta_1)$
and $m_2 = tan(\theta_2)$, we see that this is precisely what we need
for agreement with the field theory.

\newsec{A more general superpotential $W = \Tr (F \tilde F)^{k+1}$}
\subsec{Field Theory considerations}
\subseclab{\secils}

We now study the theory described in section \sectwogroups\ with a more 
general superpotential.
Let's first examine the field theory of the theory we are trying to describe 
by branes. This theory was discussed in \ils.
The gauge group is $SU(\ncl)\times SU(\ncr)$
with matter content
\thicksize=1pt
\vskip12pt
\begintable
\tstrut  | $SU(\ncl)$ | $SU(\ncr)$ | $SU(\nfl)_L$ | $SU(\nfr)_L$ | 
$SU(\nfl)_R$
 | $SU(\nfr)_R$ | $U(1)_R$ \crthick
$\Ql;\Qlt$| ${\bf\ncl;\overline\ncl}$ |${\bf 1;1}$ | 
 ${\bf \nfl;1}$ | ${\bf 1;1}$ |   ${\bf 1;\nfl}$|  ${\bf 1;1}$ 
| $1+{k\ncr - (k+1)\ncl\over\nfl (k+1)}$   \cr
$\Qr;\Qrt$|${\bf 1;1}$| ${\bf\ncr;\overline\ncr}$ |${\bf 1;1}$ | 
 ${\bf \nfr;1}$ | ${\bf 1;1}$ | ${\bf 1;\nfr}$
| $1+{k\ncl - (k+1)\ncr\over\nfr (k+1)}$   \cr
$F;\tilde F$|${\bf\ncl;\overline\ncl}$| ${\bf\ncr;\overline\ncr}$ 
|${\bf 1;1}$ |${\bf 1;1}$ | ${\bf 1;1}$ | ${\bf 1;1}$
| ${1\over k+1}$  
\endtable
\noindent
The superpotential
\eqn\Wsusu{W= \Tr(F\tilde F)^{k+1}}
truncates the chiral ring; it's equations of motion 
equate higher order operators with lower order ones. 
We studied the case $k=1$ in section \sectwogroups.
The chiral mesons
are $\Pl_{j} = \Ql (\tst\ts)^{j}\Qlt$,
$\Pr_{j} = \Qr (\tst\ts)\Qlt'$,
$\Mm_{r}=\Ql (\tst\ts)^{r-1}\tst\Qr$ and
$\Mmt_{r}=\Qlt
\ts(\tst\ts)^{r-1}\Qlt'$, where $j=0\dots k$ and
$r=0\dots k-1$.

The dual theory has gauge group $SU(\ncld)\times SU(\ncrd)$, with
$\ncld = (k+1)(\nfl+\nfr)-\nfl-\ncr$ and
$\ncrd = (k+1)(\nfr+\nfl)-\nfr-\ncl$.
The matter content is 
\thicksize=1pt
\vskip12pt
\begintable
\tstrut  | $SU(\ncld)$ | $SU(\ncrd)$ | $SU(\nfl)_L$ | $SU(\nfr)_L$ 
 | $SU(\nfl)_R$ | $SU(\nfr)_R$ | $U(1)_R$ \crthick
$\ql;\qlt$| ${\bf\ncld;\overline\ncld}$ |${\bf 1;1}$ |  ${\bf 1;1}$ |
 ${\bf \overline\nfr;1}$ | ${\bf 1;1}$ |   ${\bf 1;\overline\nfr}$ 
| $1+{k\ncrd - (k+1)\ncld\over\nfr (k+1)}$   \cr
$\qr;\qrt$|${\bf 1;1}$| ${\bf\ncrd;\overline\ncrd}$ | 
 ${\bf \overline\nfl;1}$ | ${\bf 1;1}$ | ${\bf 1;\overline\nfl}$| ${\bf 1;1}$ 
| $1+{k\ncld - (k+1)\ncrd\over\nfl (k+1)}$   \cr
$f;\tilde f$|${\bf\ncld;\overline\ncld}$
| ${\bf\ncrd;\overline\ncrd}$ 
|${\bf 1;1}$ |${\bf 1;1}$ | ${\bf 1;1}$ | ${\bf 1;1}$
| ${1\over k+1}$
\endtable
\noindent
Notice that in the dual, the fundamental fields have switched global 
symmetries.
There are also singlet fields
$\Pl_{j}$, $\Pr_{j}$,
$\Mm_{r}$, and $\Mmt_{r}$ which are a one-to-one 
map of the mesons of the electric theory.
The dual superpotential is
\eqn\WDsusu{\eqalign{
W= \Tr (f\tilde f)^{k+1}
+\sum_{j=0}^{k} \left[\Pl_{k-j} \qr (\tstD\tsD)^{j}\qrt
+\Pr_{k-j} \qlt (\tsD\tstD)^{j}\ql\right]\cr
+\sum_{r=0}^{k-1} \left[\Mm_{k-r-1} \ql \tstD(\tsD\tstD)^{r}\qr
+\Mmt_{k-r-1} \qrt (\tsD\tstD)^{r}\tsD\qlt\right].}}

Under perturbation by a mass term,
$W = \Tr(F\tilde F)^{k+1} + m \Tr F\tilde F$,
the gauge group breaks to a product of decoupled defining models
\eqn\MPsusu{SU(\ncl-\ncr+p_0)\times SU(p_0)
\times U(p_1)\times\ldots\times U(p_k)}
(for $\ncl\geq\ncr)$ with $\sum_{\ell=0}^{k}p_\ell =\ncr$.
The first (second) factor has $\nfl$ ($\nfr$) fundamental
flavors,  while the others have $\nfl+\nfr$ flavors.
The magnetic theory flows to the dual of this product.

If $\nfr+\ncl=\ncr + 1$ the $SU(\ncr)$ factor
can confine \nati\ leaving a theory of \kut, \Ahsonyank\ (see sect.
\secKEG
) with $SU(\ncl)$, an adjoint tensor $\widehat X\sim
F\tilde F$, $\nfl+\nfr + 1$ flavors and
$W=\Tr \widehat X^{k+1}$.   In the magnetic theory
$SU(\ncrd)$ confines similarly (since $\nfr+\ncld=\ncrd+1$)
leaving a theory with $SU(\tilde N_c)=SU[k(\nfl+\nfr)+1-\ncl]$
and a similar superpotential;
this is dual \kut\ to the confined
electric model. 

\subsec{Brane configuration}

We will use the notation of \sectwogroups\ to label the 5-branes.
In brane language we claim that the
theory of Intriligator, Leigh, and Strassler described in section 
\secils\ corresponds to, from left to right, 
$k$ 5-branes in the orientation of $A$ 
connected to a single 
$B$ 5-brane by $N_c$ 4-branes. The single $B$ 5-brane is in turn connected 
to $k$ more $C$ 5-branes on the right by $N_c'$ 4-branes.
$N_f$ and $N_f'$ 6-branes intersect the $N_c$ and $N_c'$ 
4-branes, respectively.

We obtain the dual configuration by reversing the order of the 
5-branes and 6-branes while preserving their linking numbers.
 From left to right, we have $N_f'$ 6-branes
connected to $k$ $C$ 5-branes by $(k+1)N_f'$ 4-branes. 
The $k$ $C$ 5-branes
are connected to the single $B$ 5-brane by $\tilde N_c$ 4-branes. 
$\tilde N_c'$ 4-branes connect the single $B$ 5-branes to 
$k$ $A$ 5-branes. $(k+1)N_f$ 4-branes in turn connect the 
$k$ $A$ 5-branes to $N_f$ 6-branes. The linking numbers show that 
the number of 4-branes 
is $\tilde N_c = (k+1)(N_f + N_f') - N_f - N_c'$ and 
$\tilde N_c' = (k+1)(N_f + N_f') - N_f' - N_c$ in agreement with the 
field theory.

Moving the $k$ 5-branes apart separates the 4-branes
and breaks the gauge groups. This corresponds in field theory
to giving the superpotential
a mass $W =\Tr  (F\tilde F)^{k+1} + m\Tr F\tilde F$. 
Similarly to section \secmass, 
we see that we get the 
same gauge groups as we found in the field theory under this 
perturbation to the superpotential.

We can consider what happens if we move the single $B$ 5-brane past 
$k$ $A$ 5-branes. This constitutes a Seiberg duality since we have moved 
a single 5-brane through $N_f$ 6-branes and avoided $k$ $A$ 5-branes 
in $x^7$.  
If $N_f' + N_c = N_c'+1$,
then there will be one 4-branes between the $B$ and $A$ 5-branes; 
the gauge group has confined. 
The brane configuration between the $A$ and $C$ 5-branes 
is that of \keg\ reviewed
in section \secKEG. We have realized in terms of branes the confinement 
described in the 
field theory in section \secils.

\newsec{A product of two gauge groups with adjoint matter.}
\subsec{Field theory considerations}
\subseclab{\secjb}

We can now use what we have learned about branes 
to understand another theory in terms of branes 
that was originally formulated in terms of field theory in 
\brodie\ and \bs. The gauge groups is
$SU(N_c) \times SU(N'_c)$ with matter 
content
\thicksize=1pt
\vskip12pt
\begintable
\tstrut  | $SU(\ncl)$ | $SU(\ncr)$ | $SU(\nfl)_L$ | $SU(\nfr)_L$ | 
$SU(\nfl)_R$
 | $SU(\nfr)_R$ | $U(1)_R$ \crthick
$\Ql;\Qlt$| ${\bf\ncl;\overline\ncl}$ |${\bf 1;1}$ | 
 ${\bf \nfl;1}$ | ${\bf 1;1}$ |   ${\bf 1;\nfl}$|  ${\bf 1;1}$ 
| $1+{\ncr - 2\ncl\over\nfl (k+1)}$   \cr
$\Qr;\Qrt$|${\bf 1;1}$| ${\bf\ncr;\overline\ncr}$ |${\bf 1;1}$ | 
 ${\bf \nfr;1}$ | ${\bf 1;1}$ | ${\bf 1;\nfr}$
| $1+{\ncl - 2\ncr\over\nfr (k+1)}$   \cr
$F;\tilde F$|${\bf\ncl;\overline\ncl}$| ${\bf\ncr;\overline\ncr}$ 
|${\bf 1;1}$ |${\bf 1;1}$ | ${\bf 1;1}$ | ${\bf 1;1}$
| ${k\over k+1}$   \cr
$X_1$ | ${\bf\ncl^2 -1}$ | ${\bf 1}$ | ${\bf 1}$ | ${\bf 1}$| ${\bf 1}$
|  ${\bf 1}$ |  ${2\over k+1}$    \cr
$X_2$ |${\bf 1}$|${\bf\ncr^2 -1}$ | ${\bf 1}$ | ${\bf 1}$ | ${\bf 1}$ | 
${\bf 1}$ |   ${2\over k+1}$
\endtable
\noindent
The superpotential is 
\eqn\AsupND{W = \Tr X_1^{k+1}
+ \Tr X_2^{k+1}
+ \Tr X_1\tilde F F - \Tr X_2\tilde F F + \rho_1\Tr  X_1
+ \rho_2 \Tr X_2. }
$\rho_1$ and $\rho_2$ are Lagrange multipliers to enforce the 
tracelessness condition.
Here $k$ can be any positive integer. It follows from the conditions for a 
supersymmetric vacuum
that the chiral ring truncates.
The gauge invariant mesons in the theory are 
$QX_1^j\tilde Q$, 
$Q'X_2^j\tilde Q'$,
$QX_1^j\tilde F Q'$,
$\tilde Q F X_2^j\tilde Q'$,
$Q\tilde F F X_1^j\tilde Q$,
$Q'\tilde FFX_2^j\tilde Q'$,
where $j=0\cdots k-1$.

The dual theory is described by an
$SU(2kN_f' + kN_f - N_c') \times SU(2kN_f + kN_f' - N_c)$ gauge theory with
matter content
\thicksize=1pt
\vskip12pt
\begintable
\tstrut  | $SU(\ncld)$ | $SU(\ncrd)$ | $SU(\nfl)_L$ | $SU(\nfr)_L$ 
 | $SU(\nfl)_R$ | $SU(\nfr)_R$ | $U(1)_R$ \crthick
$\ql;\qlt$| ${\bf\ncld;\overline\ncld}$ |${\bf 1;1}$ |  ${\bf 1;1}$ |
 ${\bf \overline\nfr;1}$ | ${\bf 1;1}$ |   ${\bf 1;\overline\nfr}$ 
| $1+{\ncrd - 2\ncld\over\nfr (k+1)}$   \cr
$\qr;\qrt$|${\bf 1;1}$| ${\bf\ncrd;\overline\ncrd}$ | 
 ${\bf \overline\nfl;1}$ | ${\bf 1;1}$ | ${\bf 1;\overline\nfl}$| ${\bf 1;1}$ 
| $1+{\ncld - 2\ncrd\over\nfl (k+1)}$   \cr
$\bar F;\tilde{\bar F}$|${\bf\ncld;\overline\ncld}$
| ${\bf\ncrd;\overline\ncrd}$ 
|${\bf 1;1}$ |${\bf 1;1}$ | ${\bf 1;1}$ | ${\bf 1;1}$
| ${k\over k+1}$   \cr
$\bar X_1$ | ${\bf\ncld^2 -1}$ | ${\bf 1}$ | ${\bf 1}$ | ${\bf 1}$| ${\bf 1}$
|  ${\bf 1}$ |  ${2\over k+1}$    \cr
$\bar X_2$ |${\bf 1}$|${\bf(\ncrd)^2 -1}$ | ${\bf 1}$ | ${\bf 1}$ 
| ${\bf 1}$ | ${\bf 1}$ |   ${2\over k+1}$
\endtable
\noindent
There are also gauge singlet mesons in the magnetic theory which
are the images of mesons in the electric theory \brodie. The dual
superpotential is analogous to \AsupND\ with the addition of coupling terms
between singlets and dual mesons.
For the case $k=1$, this reduces to the model discussed in 
section \sectwogroups.

When $kN_f + N_c' = N_c+1$ the $SU(N_c)$ theory confines. 
Fields $F\tilde F$ becomes one field $\hat Y$, and
in the superpotential we have three adjoints fields 
$X_1$, $X_2$, and $\hat Y$
where only one linear combination of $V = X_1 + X_2$ is massless. 
The magnetic theory confines similarly (since 
$kN_f' + \tilde N_c' = \tilde N_c+1$) leaving 
$SU(k(N_f + N_f') + 1- N_c')$ and $W = V^{k+1}$. The theory
therefore confines to a Kutasov-type theory with $N_f + N_f'+1$ flavors 
\kut, \Ahsonyank.

\subsec{Brane configuration}

Here we use $A$, $B$, and $C$ to denote 5-branes with orientations
as discussed in section \sectwogroups.
In brane language we claim that the model discussed in 
section \secjb\ corresponds to, from left to right, 
$k$ $A$ 5-branes connected to $k$ 
$B$ 5-branes by $N_c$ D4-branes. The $k$ $B$ 5-branes are in turn connected 
to $k$ $C$ 5-brane by $N_c'$ 4-branes. The adjoint
fields $X_1$ and $X_2$  correspond to the motion of the $k$ 5-branes
in $(x^4,x^5,x^8,x^9)$.
$N_f$ and $N_f'$ 6-branes intersect the $N_c$ and $N_c'$ 
4-branes respectively.

The dual configuration for this theory is $k$ copies of the theory 
discussed in section \sectwogroups. Linking numbers 
can be used to show that the dual gauge group is 
$SU(2kN_f + kN_f' - N_c)\times SU(2kN_f' + kN_f - N_c')$.

If we switch all of the $k$ $A$ 5-branes with the $k$ $B$ 5-branes,
and we satisfy the condition $kN_f + N_c' = N_c+1$, then we will have one 
4-brane between the $B$ and $A$ 5-branes and $N_c'$ 4-branes between the 
$A$ and $C$ 5-branes. 
Now, if we reverse the order of the $A$ and $C$ branes,
the linking numbers give 
$kN_f + kN_f' + 1 - N_c'$ 4-branes.
The $SU(N_c)$ group has confined, and 
we are left with the brane configuration of the Kutasov and Schwimmer
duality discussed in section \secKEG\ and in \keg\ in agreement 
with the field theory discussed in section \secjb.

\newsec{More than two groups dualities}

Given that we have been able to form product dualities 
by generalizing a brane configuration involving one set of 4-branes 
suspended between two 5-branes to a 
configuration involving two sets of 4-branes suspended between 
three NS 5-branes, it is natural to ask 
if we can generalize this to configurations with three or more 
sets of 4-branes 
suspended between four or more NS 5-branes. We find that the
answer is yes; such duals do exist. The field theory corresponding to such 
brane configurations have not been analyzed prior to this paper.

\subsec{$SU(N_c)\times SU(N_c')\times SU(N_c'')$}
\subseclab{\secthreegroups}

Consider 
$SU(N_c)\times SU(N_c')\times SU(N_c'')$ with a field $F$ charged under 
$\bf {(N_c,\bar N_c')}$ and a field $G$ charged under $\bf {(N_c',\bar N_c'')}$
and their conjugates $\tilde F$ and $\tilde G$. 
There are also
$N_f$ fields $Q$, $N_f'$ fields $Q'$, $N_f''$ fields $Q''$ and their 
conjugates, $\tilde Q$, $\tilde Q'$, and $\tilde Q''$, that 
transform in the fundamental representation of their 
respective gauge groups (where we have suppressed all indices). 

\def\nfl{{N_{f}}}\def\nfm{{N'_{f}}}\def\nfr{{N''_{f}}}
\def\ncl{{\nc}}\def\ncm{{N'_{c}}}\def\ncr{{N''_{c}}}

\def\ncld{{\tilde{N}_c}}\def\ncmd{{\tilde{N}'_c}}
\def\ncrd{{\tilde{N}''_c}}

\thicksize=1pt
\vskip12pt
\begintable
\tstrut  | $Q;\tilde Q$ | $Q';\tilde Q'$ |$Q'';\tilde Q''$ 
| $F;\tilde F$ |$G;\tilde G$ 
     \crthick
$SU(\ncl)$| ${\bf\ncl;\overline\ncl}$ |${\bf 1;1}$ | 
 ${\bf 1;1}$ |${\bf\ncl;\overline\ncl}$  |   ${\bf 1;1}$    \cr
$SU(\ncm)$|${\bf 1;1}$| ${\bf\ncm;\overline\ncm}$ |${\bf 1;1}$ | 
 ${\bf\overline\ncm;\ncm}$ |${\bf\ncm;\overline\ncm}$ \cr
$SU(\ncr)$|${\bf 1;1}$| ${\bf 1;1}$ 
|${\bf\ncr;\overline\ncr}$ |${\bf 1;1}$  |${\bf\overline\ncr;\ncr}$   \cr
$SU(\nfl)_R$| ${\bf 1;\nfl}$  | ${\bf 1;1}$ | ${\bf 1;1}$ | ${\bf 1;1}$
|  ${\bf 1;1}$    \cr
$SU(\nfl)_L$| ${\bf\overline\nfl;1}$ |  ${\bf 1;1}$ |  ${\bf 1;1}$ | ${\bf 1;1}$
|${\bf 1;1}$ \cr
$SU(\nfm)_R$|  ${\bf 1;1}$ | ${\bf 1;\nfm}$ |  ${\bf 1;1}$ | ${\bf 1;1}$
|${\bf 1;1}$     \cr
$SU(\nfm)_L$|  ${\bf 1;1}$ |${\bf \overline\nfm;1}$  | ${\bf 1;1}$ | ${\bf 1;1}$ 
|${\bf 1;1}$     \cr
$SU(\nfr)_R$|  ${\bf 1;1}$ | ${\bf 1;1}$ | ${\bf 1;\nfr}$ | ${\bf 1;1}$
|${\bf 1;1}$     \cr
$SU(\nfr)_L$|  ${\bf 1;1}$ | ${\bf 1;1}$ | ${\bf \overline\nfr;1}$ 
| ${\bf 1;1}$|${\bf 1;1}$     \cr
$U(1)_R$ |$1+{\ncm-2\ncl\over 2\nfl}$|$1+{\ncr+\ncl-2\ncm\over 2\nfm}$ 
| $1+{\ncm-2\ncr\over 2\nfr}$ | ${1\over 2}$ | ${1\over 2}$  
\endtable
\noindent
To truncate the chiral ring we add the
superpotential
\eqn\threegroups{W = {1\over 2}\Tr (F \tilde F)^2 + \Tr F \tilde F G \tilde G 
- {1\over 2}\Tr (G \tilde G)^2.}
The equations determining the supersymmetric minima are
\eqn\min{
\eqalign{
\tilde F F \tilde F + \tilde F G \tilde G = 0\cr
\tilde G F \tilde F - \tilde G G \tilde G = 0\cr
F \tilde F F + G \tilde G F = 0\cr
F \tilde F G - G \tilde G G = 0\cr}}
Multiplying the first equation in \min\ by $\tilde F F$ from the left 
and the right, we obtain 
\eqn\minmore{
\eqalign{
\tilde F F \tilde F F \tilde F = - \tilde F F \tilde F G \tilde G = 
\tilde F G \tilde G G \tilde G \cr
\tilde F F \tilde F F \tilde F = - \tilde F G \tilde G F \tilde F = 
- \tilde F G \tilde G G \tilde G \cr}}
where we have used the first equation in \min\ in the first equation in 
\minmore\ and the second equation in \min\ in the second equation in 
\minmore. Thus, 
$\tilde F F \tilde F F \tilde F = \tilde G G \tilde G G \tilde G = 0$. 
and $\tilde F F \tilde F F = -\tilde G G \tilde G G$.

The mesons in the theory are 
$M^{0,0}_0 = Q\tilde Q$, 
$M^{0,0}_2 = Q\tilde F F\tilde Q$, 
$M^{0,0}_4 = Q\tilde F F\tilde F F \tilde Q$, 
$M^{0,1}_1 = Q\tilde F Q'$, 
$M^{0,1}_3 = Q\tilde F F\tilde F Q'$,
$M^{0,2}_2 = Q\tilde F G\tilde Q''$, 
$\tilde M^{0,1}_1 = \tilde Q F \tilde Q'$, 
$\tilde M^{0,1}_3 = \tilde Q F \tilde F F \tilde Q'$,
$\tilde M^{0,2}_2 = \tilde Q F \tilde G Q''$, 
$M^{1,1}_0 =  Q'\tilde Q'$, 
$M^{1,1}_{2_F} = Q'\tilde F F\tilde Q'$,
$M^{1,1}_{2_G} = Q'\tilde G G\tilde Q'$, 
$M^{1,1}_4 = Q'\tilde F F\tilde F F \tilde Q'$, 
$M^{2,2}_0 = Q''\tilde Q''$, 
$M^{2,2}_2 = Q''\tilde G G\tilde Q''$, 
$M^{2,2}_4 = Q''\tilde G G\tilde G G \tilde Q''$,
$M^{1,2}_1 = Q'\tilde G Q''$ 
$M^{1,2}_3 = Q'\tilde G G\tilde G Q''$
$\tilde M^{1,2}_1 = \tilde Q' G \tilde Q''$ 
and
$\tilde M^{1,2}_3 = \tilde Q' G \tilde G G \tilde Q''$. 

The dual gauge group is $SU(3N_f'' + 2N_f' + Nf - N_c'')
\times SU(4N_f' + 2N_f + 2N_f'' - N_c')\times SU(3N_f + 2N_f' + N_f'' - N_c)$
 with a field $f$ charged under 
$\bf {(\tilde N_c,\bar {\tilde N_c'})}$ 
and a field $g$ charged under $(\tilde N_c',\bar {\tilde N_c''})$
and their conjugates $\tilde f$ and $\tilde g$. 
There are also
$N_f''$ fields $q$, $N_f'$ fields $q'$, $N_f$ fields $q''$ and their 
conjugates, $\tilde q$, $\tilde q'$, and $\tilde q''$, that 
transform in the fundamental representation of 
$SU(\tilde N_c)\times SU(\tilde N_c') \times SU(\tilde N_c'')$ 
respectively.
We note that as in section \secils, the fundamentals 
have reversed their global symmetries in the dual. 
\thicksize=1pt
\vskip12pt
\begintable
\tstrut  | $q;\tilde q$ | $q';\tilde q'$ |$q'';\tilde q''$ 
| $f;\tilde f$ |$g;\tilde g$ 
     \crthick
$SU(\ncld)$| ${\bf\ncld;\overline\ncld}$ |${\bf 1;1}$ | 
 ${\bf 1;1}$ |${\bf\ncld;\overline\ncld}$  |   ${\bf 1;1}$    \cr
$SU(\ncmd)$|${\bf 1;1}$| ${\bf\ncmd;\overline\ncmd}$ |${\bf 1;1}$ | 
 ${\bf\overline\ncmd;\ncmd}$ |${\bf\ncmd;\overline\ncmd}$ \cr
$SU(\ncrd)$|${\bf 1;1}$| ${\bf 1;1}$ 
|${\bf\ncrd;\overline\ncrd}$ |${\bf 1;1}$  |${\bf\overline\ncrd;\ncrd}$   \cr
$SU(\nfl)_R$| ${\bf 1;1}$ | ${\bf 1;1}$ | ${\bf 1;\nfl}$  | ${\bf 1;1}$
|  ${\bf 1;1}$    \cr
$SU(\nfl)_L$| ${\bf 1;1}$ |  ${\bf 1;1}$ |${\bf\overline\nfl;1}$| ${\bf 1;1}$
|${\bf 1;1}$ \cr
$SU(\nfm)_R$|  ${\bf 1;1}$ | ${\bf 1;\nfm}$ |  ${\bf 1;1}$ | ${\bf 1;1}$
|${\bf 1;1}$     \cr
$SU(\nfm)_L$|  ${\bf 1;1}$ |${\bf \overline\nfm;1}$  | ${\bf 1;1}$ | ${\bf 1;1}$ 
|${\bf 1;1}$     \cr
$SU(\nfr)_R$|${\bf 1;\nfr}$ |  ${\bf 1;1}$ | ${\bf 1;1}$ | ${\bf 1;1}$
|${\bf 1;1}$     \cr
$SU(\nfr)_L$ | ${\bf \overline\nfr;1}$ 
|  ${\bf 1;1}$ |  ${\bf 1;1}$|${\bf 1;1}$ |${\bf 1;1}$    \cr
$U(1)_R$ |$1+{2\ncr-\ncl-4\nfr\over 2\nfr}$|$1+{2\ncm-\ncl-\ncr-4\nfm\over 
2\nfm}$ 
| $1+{2\ncl-\ncm-4\nfl\over 2\nfl}$ | ${1\over 2}$ | ${1\over 2}$  
\endtable
\noindent
There is a one-to-one map from the mesons of the original theory to 
the singlets of the dual theory. The dual superpotential is 
analogous to \threegroups, but has coupling to singlets.
\eqn\dualW{
\eqalign{
W = & \Tr (f \tilde f)^2 + \Tr f \tilde f g\tilde g - \Tr g \tilde g g \tilde g 
+ M^{0,0}_0\tilde q'' f\tilde f f\tilde f q'' 
+ M^{0,0}_2\tilde q''  f \tilde f q'' + M^{0,0}_4\tilde q'' q'' \cr
+&M^{0,1}_1\tilde q' f \tilde f f \tilde q'' 
+ M^{0,1}_3\tilde q' f \tilde q'' 
+ \tilde M^{0,1}_1q' \tilde f f \tilde f q'' 
+ \tilde M^{0,1}_3q' \tilde f q''  
+ M^{0,2}_2\tilde q f \tilde g q'' 
+ \tilde M^{0,2}_2q \tilde f g \tilde q'' \cr
+&M^{1,1}_0\tilde q' f\tilde f f\tilde f q'  
+ M^{1,1}_{2_G}\tilde q' f \tilde f q'
+ M^{1,1}_{2_F}\tilde q' g \tilde g q'  
+ M^{1,1}_4\tilde q' q' \cr
+&M^{2,2}_0\tilde q f\tilde f f\tilde f q 
+ M^{2,2}_2\tilde q  f \tilde f q 
+ M^{2,2}_4\tilde q q \cr
+&M^{1,2}_1\tilde q' f \tilde f f \tilde q 
+ M^{1,2}_3\tilde q' f \tilde q
+ \tilde M^{1,2}_1q' \tilde f f \tilde f q 
+ \tilde M^{1,2}_3 q' \tilde f q. \cr}}
The fact that the dual superpotential is consistent 
with the global symmetries is a non-trivial test of the 
duality.
We have checked that the 't Hooft anomaly matching conditions are 
satisfied.

On the electric side,
it is possible to add a mass term for $F$. 
\eqn\threegroupsmass{W = \Tr (F \tilde F)^2 + \Tr F \tilde F G \tilde G 
- \Tr (G \tilde G)^2 + m\Tr F\tilde F.}
The supersymmetric minima allows for a generic solution where $<F\tilde F>$
and $<G\tilde G>$ are not zero, and the gauge
group is broken to $SU(N_c - r_1 - r_2)$ with $N_f$ flavors, 
$SU(N_c' - r_1 - r_2 - r_3)$ with $N_f'$ flavors,
$SU(N_c'' - r_2 - r_3)$ with $N_f''$ flavors, 
$U(r_1)$ with $N_f + N_f'$ flavors, 
$U(r_3)$ with $N_f' + N_f''$ flavors, and $U(r_2)$ with $N_f + N_f' + N_f''$ 
flavors. The fields $F$ and $G$ are generically massive. In the dual theory,
we add the dual operator $mf\tilde f$ to the dual superpotential. 
The dual theory breaks to 
$SU(N_f - N_c + r_1 + r_2)$ with $N_f$ flavors, 
$SU(N_f' - N_c' + r_1 + r_2 + r_3)$ with $N_f'$ flavors,
$SU(N_f'' - N_c'' + r_2 + r_3)$ with $N_f''$ flavors, 
$U(N_f + N_f' - r_1)$ with $N_f + N_f'$ flavors, 
$U(N_f' + N_f'' - r_3)$ with $N_f' + N_f''$ flavors, and 
$U(N_f + N_f' + N_f'' - r_2)$ with $N_f + N_f' + N_f''$ 
flavors.

There is a special
case where $r_1 = r_2 = r_3 = 0$ and $<F \tilde F> = <G\tilde G> = 0$ where 
the fields $G$ and $\tilde G$ are massless. The theory then flows to a 
Seiberg duality and the duality discussed in section \sectwogroups\
and in \ils.
The fact that we can connect this duality to other known
dualities is another important consistency check.

\ifig\figthreegroups{The theory has 
$SU(N_c)\times SU(N_c')\times SU(N_c'')$
gauge fields living on the D4-branes. Strings stretching between the 
first two sets of 4-branes are the fields $F$ in the $\bf {(N_c, \bar N_c')}$ 
representation. Strings stretching between the 
second two sets of 4-branes are the fields $G$ in the $\bf 
{(N_c', \bar N_c'')}$ 
representation. There are $N_f$ strings stretching from a 4-brane to the 
$N_f$ 6-branes. These are the fundamental fields $Q$, $Q'$, and $Q''$. There
is a superpotential $W = \Tr (F \tilde F)^2 + \Tr F \tilde F G \tilde G 
- \Tr (G \tilde G)^2$.}
{\epsfxsize3.0in\epsfbox{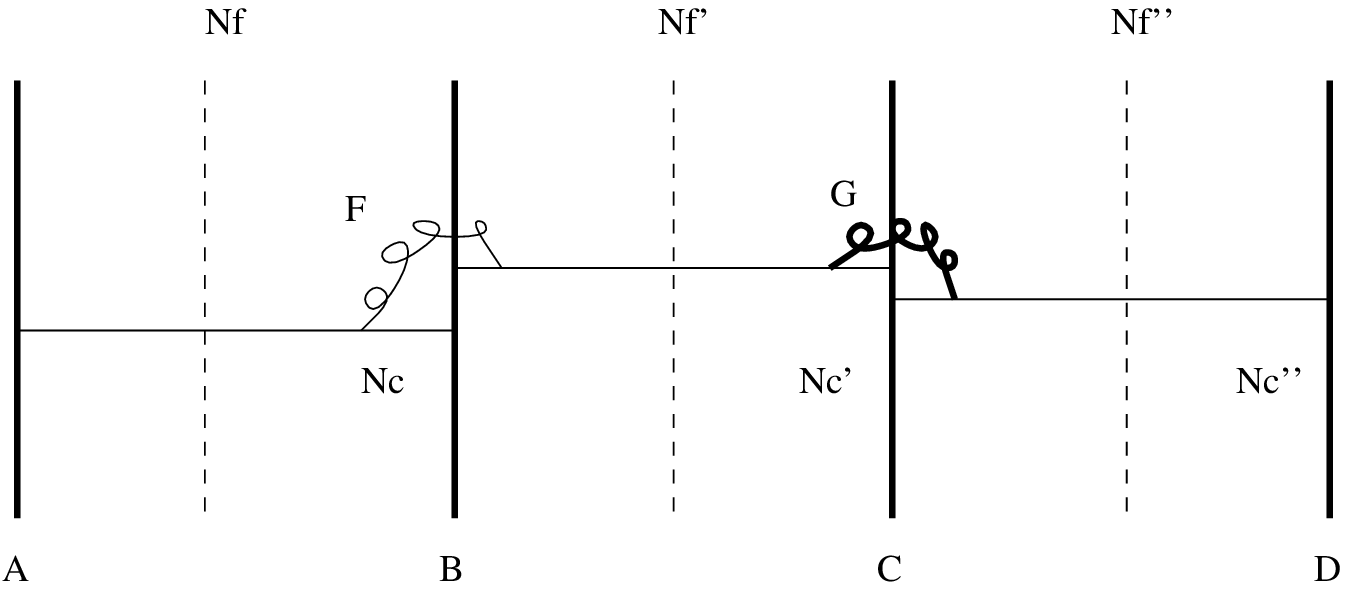}}

\subsec{Brane configurations of theories with three gauge groups}

To describe a theory with three gauge groups we must have four 
5-branes. We will label these branes
$A$, $B$, $C$, and $D$ (we hope that the reader will not find a D NS 5-brane
confusing). As before, the four 5-branes are oriented at 
different angles in $(x^4,x^5,x^8,x^9)$; 
it is important that no two 5-branes are parallel.
Consider the brane configuration described in \figthreegroups. 
 From left to right,
we have one $A$ 5-brane connected to a $B$ 5-brane 
by $N_c$ 4-branes.
The $B$ NS 5-brane is connected to a $C$ 5-brane 
by $N_c'$ 4-branes.
The $C$ 5-brane is connected to another NS 5-brane, $D$, 
by $N_c''$ 4-branes.
The gauge group is then $SU(N_c)\times SU(N_c')\times SU(N_c'')$.
$N_f$, $N_f'$ and $N_f''$ 6-branes intersect the 
$N_c$, $N_c'$, and $N_c''$ 4-branes in the $x^6$ direction.
Strings going from the $N_c$ $N_c'$, and $N_c''$ 
4-branes to the $N_f$ $N_f'$ and $N_f''$ 6-branes give us the
fields $Q$, $Q'$ and $Q''$ and 
$\tilde Q$, $\tilde Q'$, and  $\tilde Q''$ respectively. 
Strings between the set of $N_c$ 4-branes and 
the set of $N_c'$ 4-branes are the fields $F$ and $\tilde F$ 
in the representation $\bf{(N_c,\bar N_c', 1)}$ and  $\bf{(\bar N_c,N_c', 1)}$.
Strings between the set of $N_c'$ 4-branes and 
the set of $N_c''$ 4-branes are the fields $G$ and $\tilde G$ 
in the representation $\bf{(1,N_c',\bar N_c'')}$ and  
$\bf{(1,\bar N_c',N_c'')}$.
\ifig\figthreedual{Here we have 
the dual configuration 
$SU(\tilde N_c)\times SU(\tilde N_c')\times SU(\tilde N_c'')$
on the world volume of the 4-branes. Note the appearance of semi-infinite
4-branes.}
{\epsfxsize6.0in\epsfbox{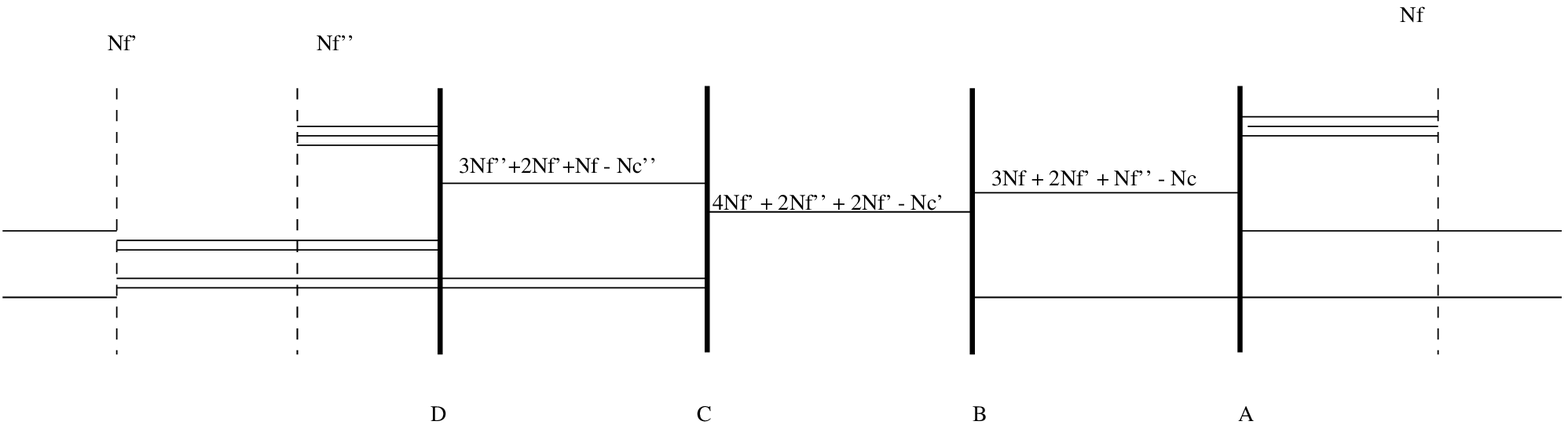}}

In the electric theory, the 5-branes are ordered $A,B,C,D$. To find the 
magnetic theory, we claim we should reverse the ordering to 
$D,C,B,A$. The 6-branes also have their ordering inverted. 
In order to have the linking numbers match up with the results of field theory
we found it necessary to introduce some semi-infinite 4-branes. The 
dual brane configuration is shown in \figthreedual. From left to right we have 
$2N_f'$ semi-infinite 4-branes ending on $N_f'$ 6-branes. 
$2N_f'$ 4-branes stretch between 
the $N_f'$ 6-branes and the NS 5-brane which has been labeled $D$. 
$2N_f'$ more 4-branes stretch between the $N_f'$ 6-branes and the $C$ 5-brane.
Proceeding to the right, $3N_f''$ 4-branes stretch between the 
$N_f''$ 
6-branes and the $D$ NS 5-brane. $3N_f'+2N_f''+N_f - N_c''$ 4-branes
stretch between the $D$ NS 5-brane and the $C$ 5-brane. 
$4N_f'+2N_f''+2N_f - N_c'$ 4-branes
stretch between the $C$ 5-brane and the $B$ 5-brane. 
$3N_f+2N_f'+N_f'' - N_c$ 4-branes
stretch between the $B$ 5-brane and the $A$ 5-brane.   
Attached to the $B$ and $A$ 5-branes are $N_f'$ semi-infinite 4-branes
each extending to the right. Finally, there are $3N_f$ 4-branes stretched 
between
the $A$ 5-brane and $Nf$ 6-branes. It is interesting to note that the 
configuration
looks more natural when $x^6$ is compact enabling 
the semi-infinite 4-branes to 
connect.

\subsec{Three product dualities with adjoints}
\subseclab{\secthreeadj}

It is also possible to form products of three gauge groups with 
adjoint matter. Consider 
$SU(N_c)\times SU(N_c')\times SU(N_c'')$ with fields $X_1$
in the adjoint representation of $SU(N_c)$, $X_2$
in the adjoint representation of $SU(N^{'}_c)$, and $X_3$
in the adjoint representation of $SU(N^{''}_c)$
As before, we also have fields
$F$ charged under 
$\bf {(N_c,\bar N_c')}$ and fields $G$ charged under $\bf {(N_c',\bar N_c'')}$
and their conjugates $\tilde F$ and $\tilde G$. 
There are also
$N_f$ fields $Q$, $N_f'$ fields $Q'$, $N_f''$ fields $Q''$ and their 
conjugates, $\tilde Q$, $\tilde Q'$, and $\tilde Q''$, that 
transform in the fundamental representation of their 
respective gauge groups. To truncate the chiral ring we add the
superpotential
\eqn\threegroupsadj{\eqalign{W = & {1\over k+1}\Tr {X_1}^{k+1} 
+ \Tr X_1 F \tilde F +  {1\over k+1}\Tr X_2^{k+1} 
+  \Tr X_2 G \tilde G +  \Tr X_2 F \tilde F \cr 
+ &{1\over k+1}\Tr {X_3}^{k+1} -  \Tr X_3 G \tilde G 
 +  \rho_1\Tr X_1 + \rho_2\Tr X_2 + \rho_3\Tr X_3.}}
The conditions for supersymmetric minima are 
\eqn\mink{
\eqalign{
X_1^k + \tilde F F + \rho_1= 0\cr
X_2^k + F \tilde F - \tilde G G +\rho_2= 0\cr
X_3^k + G \tilde G +\rho_3= 0\cr
X_1F - FX_2 = 0\cr
X_1\tilde F - \tilde FX_2 = 0\cr
X_2G - GX_3 = 0\cr
X_2\tilde G - \tilde GX_3 = 0\cr
}}
For $k$ odd it is possible to show that equations \mink\ reduce to 
equations \min, and thus the chiral ring truncates.
The mesons in the theory are 
$M^{0,0}_{0,j} = Q{X_1}^j \tilde Q$, 
$M^{0,0}_{2,j} = Q{X_1}^j \tilde F F\tilde Q$, 
$M^{0,0}_{4,j} = QX_1^j \tilde F F\tilde F F \tilde Q$, 
$M^{0,1}_{1,j} = QX_1^j \tilde F Q'$, 
$M^{0,1}_{3,j} = QX_1^j \tilde F F\tilde F Q'$,
$M^{0,2}_{2,j} = QX_1^j \tilde F G\tilde Q''$, 
$\tilde M^{0,1}_{1,j} = \tilde QFX_2^j \tilde Q'$, 
$\tilde M^{0,1}_{3,j} = \tilde QF \tilde F\tilde F \tilde X_2^j Q'$,
$\tilde M^{0,2}_{2,j} = \tilde QF \tilde G X_3^j Q''$, 
$M^{1,1}_{0,j} = Q'X_2^j \tilde Q'$, 
$M^{1,1}_{2_{F},j} = Q'X_2^j \tilde F F\tilde Q'$,
$M^{1,1}_{2_{G},j} = Q'X_2^j \tilde G G\tilde Q'$, 
$M^{1,1}_{4,j} = Q'X_2^j \tilde F F\tilde F F \tilde Q'$, 
$M^{2,2}_{0,j} = Q''X_3^j \tilde Q''$, 
$M^{2,2}_{2,j} = Q''X_3^j \tilde G G\tilde Q''$, 
$M^{2,2}_{4,j} = Q''X_3^j \tilde G G\tilde G G \tilde Q''$, 
$M^{1,2}_{3,j} = Q'X_2^j \tilde G G \tilde G Q''$
$\tilde M^{1,2}_{1,j} = \tilde Q' G X_2^j  \tilde Q''$
and 
$M^{1,2}_{1,j} = Q'X_2^j \tilde G Q''$ where $j = 0,\cdots ,k$. 

The dual gauge group is $SU(3kN_f'' + 2kN_f' + kNf - N_c'')
\times SU(4kN_f' + 2kN_f + 2kN_f'' - N_c')
\times SU(3kN_f + 2kN_f' + kN_f'' - N_c)$.
There is similar matter content charged under the dual gauge group, and 
as usual, a map from mesons of the original theory to singlets of 
the dual theory.
The dual superpotential is
\eqn\dualWk{
\eqalign{
W = & \Tr {\bar X_1}^{k+1} + \Tr \bar X_1 f \tilde f +  \Tr {\bar X_2}^{k+1} 
+  \Tr \bar X_2 f \tilde f +  \Tr \bar X_2 g \tilde g 
+  \Tr {\bar X_3}^{k+1} -  \Tr \bar X_3 g \tilde g \cr
+& \rho_1\Tr \bar X_1 + \rho_2\Tr \bar X_2 + \rho_3\Tr \bar X_3\cr
+&M^{0,0}_{0,j}\tilde q'' \bar X_3^j f\tilde f f\tilde f q'' 
+ M^{0,0}_{2,j}\tilde q'' \bar X_3^j f \tilde f q'' 
+ M^{0,0}_{4,j} \tilde q'' \bar X_3^j q'' \cr
+&M^{0,1}_{1,j} \tilde q' f \tilde f f \bar X_3^j \tilde q'' 
+ M^{0,1}_{3,j}\tilde q' f \bar X_3^j \tilde q'' 
+ M^{0,2}_{1,j}\tilde q f \tilde g\bar X_3^j q'' \cr
+&\tilde M^{0,1}_{1,j}q' \tilde f f \tilde f \bar X_3^j q'' 
+ \tilde M^{0,1}_{3,j}q' \tilde f \bar X_3^j q''  
+ \tilde M^{0,2}_{1,j}q  \tilde f g \bar X_3^j \tilde q'' \cr
+&M^{1,1}_{0,j}\tilde q' \bar X_2^j f\tilde f f\tilde f q'  
+ M^{1,1}_{{2_G},j}\tilde q'\bar X_2^j f \tilde f q'
+ M^{1,1}_{{2_F},j}\tilde q' \bar X_2^j g \tilde g q'  
+ M^{1,1}_{4,j}\tilde q'\bar X_2^j  q' \cr
+&M^{2,2}_{0,j}\tilde q \bar X_1^j f\tilde f f\tilde f q 
+ M^{2,2}_{2,j}\tilde q \bar X_1^j  f \tilde f q 
+ M^{2,2}_{4,j}\tilde q \bar X_1^j q \cr
+&M^{1,2}_{1,j}\tilde q' f \tilde f f \bar X_1^j \tilde q 
+ M^{1,2}_{3,j}\tilde q' f \bar X_1^j \tilde q
+ \tilde M^{1,2}_{1,j}q' \tilde f f \tilde f \bar X_1^j q 
+ \tilde M^{1,2}_{3,j}q' \tilde f \bar X_1^j q. \cr}}
The fact that the dual superpotential is consistent 
with the global symmetries is a non-trivial test of the 
duality.
We have checked that the 't Hooft anomaly matching conditions are 
satisfied.

\subsec{Generalization to $n$ gauge groups and $n+1$ 5-branes}
\subseclab{\secngroups}

We consider theories with $n$ gauge groups: $SU(N_{c_0})\times 
SU(N_{c_1})\times ... \times SU(N_{c_{n-1}})$. Each gauge group has
fundamental fields $N_{f_0}$, $N_{f_1}$, $\cdots$ , $N_{f_{n-1}}$. 
The fields that link the gauge groups together are 
$F_{01}$ in the $\bf{(N_{c_0},\bar N_{c_1},1,1,...,1)}$, 
$F_{12}$ in the $\bf{(1,N_{c_1},\bar N_{c_2},1,...,1)}$,
$F_{23}$ in the $\bf{(1,1,N_{c_2},\bar N_{c_3},...,1)}$,
 ....
$F_{n-2,n-1}$ in the $\bf{(1,1,1,1,...,N_{c_{n-2}},\bar N_{c_{n-1}})}$
and their conjugate fields.
The superpotential is 
\eqn\nbranes{\eqalign{W  = & 
(F_{01}\tilde F_{10})^2 + F_{01}\tilde F_{10}F_{12}\tilde F_{21}
+ (F_{12}\tilde F_{21})^2 + F_{12}\tilde F_{21}F_{23}\tilde F_{32} \cr
- & (F_{23}\tilde F_{32})^2 + ...  
+ F_{n-3,n-2}\tilde F_{n-2,n-3}F_{n-2,n-1}\tilde F_{n-1,n-2}
- (F_{n-2,n-1}\tilde F_{n-1,n-2})^2 \cr}} 
The dual gauge group is 
\eqn\nbranes{\eqalign{
& SU(nN_{f_{n-1}} + (n-1)N_{f_{n-2}} + ... + N_{f_0} - N_{c_{n-1}}) \cr
\times &
SU(p_0 - N_{f_{n-1}} + (n-1)N_{f_{n-2}} + (n-2)N_{f_{n-3}} + ... 
+ N_{f_0} - N_{c_{n-2}})\cr
...\times &
SU(p_{r-1} - N_{f_{n-1}} - 2N_{f_{n-2}} - ... - rN_{f_{n-r}} 
+ (n-r)N_{f_{n-r-1}} \cr
+ & (n-r-1)N_{f_{n-r-2}} + ... + N_{f_0} - N_{c_{n-r-1}})\cr
...
\times &
SU(nN_{f_0} + (n-1)N_{f_1} + (n-2)N_{f_2} + ... + N_{f_{n-1}} - N_{c_0})\cr}}
where $p_r$ is the rank of the $r$-th dual gauge group minus 
$N_{c_{n-r}}-1$.
It is possible to add adjoint matter fields to the 
gauge theory described above in a similar manner to that
discussed in section \secthreeadj.

\newsec{Conclusions.}
D-branes constructions of type IIA string theory
have allowed us to explore four dimensional quantum field theories. 
We see familiar field theory phenomena, such as 
chiral symmetry and duality, 
in a new light. 
We have shown that in a number of cases it is 
easy to compute the dual gauge group of a theory 
by reversing the ordering of the 5-branes and 6-branes 
while preserving the linking numbers.
Branes have shown to 
be a powerful tool in guiding us to interesting, 
new field theory dualities. 
In section 6, we saw that in order to have 
linking numbers and field theory agree, we 
had to introduce semi-infinte 4-branes.
It would be interesting to understand this transition
better. 
In section \secchiralsym, we saw that incorporating 
chiral symmetry in the brane picture led us to 
predict that quantization of tensionless threebranes in six
dimensions gives rise to massless vector multiplet in six dimensions.
It would be very interesting to provide further support for this scenario.
We expect much 
progress in the near future.

\newsec{Acknowledgements}
We would like to thank Ofer Aharony, Aki Hashimoto, Peter Mayr, Ronen Plesser, 
Sanjaye Ramgoolam, Nathan Seiberg, Edward Witten, and especially 
Ken Intriligator for helpful discussions. 

\listrefs

\end